\documentclass{article}%
\usepackage{amssymb}
\usepackage{graphicx}
\usepackage{amsmath}
\usepackage{amsfonts}%
\begin{document}

\title{A multivariate interlace polynomial}
\author{Bruno Courcelle\\Bordeaux University and CNRS, LaBRI}
\date{December 2006 (revised March 2007)}
\maketitle

\begin{abstract}
We define a multivariate polynomial that generalizes several \emph{interlace}
polynomials defined by Arratia, Bollobas and Sorkin on the one hand, and
Aigner and van der Holst on the other. We follow the route traced by Sokal,
who defined a multivariate generalization of Tutte's polynomial. We also show
that bounded portions of our interlace polynomial can be evaluated in
polynomial time for graphs of bounded clique-width. Its full evaluation\ is
necessarily exponential just because of the size of the result.

\end{abstract}


Support\footnote{This work has been supported by the ANR\ project GRAAL\ and
by a temporary position of CNRS\ researcher.}

\bigskip

\section{Introduction}

\bigskip

Many polynomials associated with graphs, matroids or combinatorial maps count
\emph{configurations} in these objects.\ We take here "configuration" in a
wide sense.\ Typical examples are colorings, matchings, stable subsets,
subgraphs. In many cases, a $weight$ is associated with the considered
configurations : number of colors, cardinality, number of connected components
or rank of an associated subgraph.

A \emph{multivariate polynomial}, as the one defined by A.\ Sokal, that
generalizes Tutte's two variable polynomial, not only counts configurations,
but also enumerates them, together with their associated weights. Multivariate
polynomials may have recursive definitions, the specializations of which give
the classical recursive definitions for the corresponding polynomials.
(Specialization is the inverse of generalization, these notions are explained below).

We think that a recursive definition at the multivariate level makes better
understand what is going on at the usual level of one or two variable
polynomials. Furthermore, multivariate polynomials may have \emph{static
definitions} based on properties of configurations and weights expressible in
\emph{second-order logic} or better in \emph{monadic second-order logic}. Why
better ?\ Because this yields fixed parameter tractable algorithms, where the
parameters are the tree-width or the clique-width of a graph, or the
branch-width of a matroid.\ This consequence has been explained in [CMR,
Mak04, Mak05]. We apply these ideas to the \emph{interlace polynomials}
defined first under this name in [ABS], then generalized in [ABS04b] and [AvH]
(and previously defined under another name, see \ Las Vergnas [LV]), that is,
we give a common multivariate generalization and show that is is definable by
monadic second-order formulas.

What do we mean by "generalize" ? If two polynomials $P(G)$ and $Q(G)$ are
associated with each graph $G$ of a certain type, we say that $P(G)$ is
\emph{more general} that $Q(G)$ and that $Q(G)$ is a \emph{specialization} of
$P(G)$ if $Q(G)$ can be obtained from $P(G)$ by a \emph{substitution} of fixed
polynomials to the variables of $P(G);$ such a substitution may be also
combined with multiplication by polynomials depending, say, on the number of
vertices and/or edges. We do not try to propose here a most general
definition.\ Applications of the idea will suffice.

\bigskip

A \emph{multivariate polynomial} is one with indeterminates depending on the
vertices or the edges of the considered graph (such indeterminates are
sometimes called "weights", because they make possible to evaluate the
polynomial with distinct values associated with distinct vertices or edges).
Sokal's multivariate Tutte polynomial of a graph $G=(V,E)$ is defined by :

\begin{center}
$Z(G)=\sum\nolimits_{A\subseteq E}u^{k(G[A])}\prod\nolimits_{e\in A}v_{e}$ \ 
\end{center}

where $G[A]$ is the subgraph of $G$ with set of vertices $V$ and set of edges
$A$, $k(G[A])$ is the number of its connected components.\ This polynomial
belongs to $\mathbf{Z}[u,v_{e}$ ; $e\in E]$. An indeterminate $v_{e}$ is
associated with each edge $e$.\ The indeterminates commute, the order of
enumeration over each set $A$ is irrelevant. For two graphs $G$ and
$G^{\prime}$ with sets of edges in bijection, we have $Z(G)=Z(G^{\prime})$
(where the variables indexed by edges of $G$\ \ and $G^{\prime}$\ correspond
via the considered bijection) \ iff \ $\mid V(G)\mid=\mid V(G^{\prime})\mid$
and their cycle matroids are isomorphic (via the same bijection between
edges). This observation explains what information is contained in this
polynomial about the considered graph.

The polynomial $Z(G)$ is more general than Tutte's two variable polynomial
$T(G,x,y)$ because (see [Sok] for details) :

\begin{center}
$T(G,x,y)=((x-1)^{k(G)}(y-1)^{\mid V\mid})^{-1}\alpha(Z(G))$
\end{center}

where $\alpha$ \ is the substitution : \ 

\begin{center}
$[u:=(x-1)(y-1);v_{e}:=y-1$ \ \ for all $\ e\in E].$
\end{center}

Conversely, one can express $Z^{\prime}(G),$ defined as $Z(G)$ where every
indeterminate $v_{e}$ replaced by the same indeterminate $v,$ in terms of
$T(G,x,y)$ in a similar way. Hence, $Z^{\prime}(G)$ \ and $T(G)$ are
equivalent both in expressive power and for the complexity of their computations.

\bigskip

In this article, we define a multivariate polynomial, that generalizes the
different \emph{interlace} polynomials defined for graphs in [ABS], [ABS04b]
and [AvH], and also, the \emph{independence} polynomial surveyed in [LM]. In
our polynomial, a configuration is a set\ $A$ of vertices and its associated
value is the rank of the induced subgraph. By \emph{rank} we mean, as in these
articles, the rank of the adjacency matrix with respect to the field GF(2$).$
Actually, in order to generalize a polynomial defined in [AvH], we introduce a
second argument in a configuration, namely a set $B$ of vertices corresponding
to "toggled loops" : if a vertex in $B$ has no loop we add one ; if it has a
loop, we delete it. Then we evaluate ranks.

We find for this polynomial a recursive definition, somewhat more complicated
than the usual ones based on contracting and deleting edges. The polynomials
of [ABS], [ABS04b] and [AvH] are specializations of ours, and we find their
recursive definitions as the corresponding specializations, sometimes with the
necessity of proving nontrivial properties.

\bigskip

This approach is based on \emph{static definitions} from which \emph{recursive
definitions} can be constructed, and not the other way around.\ Here we follow
Sokal who considers recursive definitions as secundary.\ It is developped in a
fundamental perspective in [CGM]. Let us say to have a representative picture
of the most general case that the polynomials we obtain in this way are of the
form :

\begin{center}
$P(G)=\sum\nolimits_{C\in\Gamma(G)}n_{C}\cdot v_{C}u^{f(C)}$ \ 
\end{center}

where $C$ ranges over all configurations of a multiset $\Gamma(G),$ $n_{C}%
$\ is the number of occurrences of $C$ in $\Gamma(G)$, $v_{C}$ is a monomial
(like \ \ $\prod\nolimits_{e\in A}v_{e}$ in the above polynomial $Z(G)$) that
describes \ configuration $C$, and $f(C)$ is the weight of $C$. Polynomials of
this form have necessarily positive coeffficients. We are especially
interested in cases where $\Gamma(G)$ and $f$ can be expressed by monadic
second-order formulas because of algorithmic applications developped in [CMR,
Mak04, Mak05].

\bigskip

Such polynomials are usually of exponential size.\ We may ask for a polynomial
evaluation of the \emph{truncation} of $P(G)$ i.e., of its part limited to the
monomials associated with configurations of a given size.\ For such purpose,
having a description of $\Gamma(G)$ and of the weight function $f$ \ by
monadic second order formulas is helpful, as we will see in the last section,
because this yields \emph{fixed parameter tractable algorithms}, where the
parameter is tree-width or clique-width.

Summary of the article : 2.\ Definitions and basic facts, 3. A multivariate
interlace polynomial and its recursive definition, 4.\ Specializations to
known polynomials, 5. Polynomial time evaluation for graphs of bounded
clique-width, 6. Conclusion, 7.\ References, 8. Appendix : The Tutte polynomial

\bigskip

\section{Definitions and basic facts}

\bigskip

Graphs are finite, simple, undirected, possibly with loops. A graph is defined
as a pair $G=(V_{G},A_{G})$ of a set of vertices $V_{G}$ and a symmetric
adjacency matrix $\ A_{G}$ over GF(2). We omit the subscripts whenever
possible without ambiguity. The \emph{rank} $rk(G)$ of $G=(V,A)$ is defined as
the rank $rk(A)$ of $A$ over GF(2); its \emph{corank (or nullity)} is
$n(G):=n(A):=\mid V\mid-rk(A)$. The empty graph $\varnothing$\ has rank and
corank 0.

\bigskip

The set of looped vertices of $G$ (the vertices $i$ such that $A(i,i)=1)$ is
denoted by $Loops(G)$. For $a$ in $V$, we let $N(G,a)$ be the set of neighours
$b$ of $a$, with $b\neq a$. (A looped vertex is not a neighbour of itself).

If $X$ is a set of vertices, we let $G-X$ denote $G[V-X]$, the induced
subgraph of $G$\ with set of vertices $V-X$.

We denote by $G\nabla X$ the graph obtained by "toggling" the loops in $X$,
i.e., \ $V_{G\nabla X}:=V_{G}$ and \ :

$\qquad A_{G\nabla X}(i,j):=1-A_{G}(i,j)$ \ if \ $i=j\in X,$

$\qquad A_{G\nabla X}(i,j):=A_{G}(i,j)$ \ otherwise.

\bigskip

We write $G=H\oplus K$\ if $G$ is the union of disjoint subgraphs $H$ and $K$.

For two graphs $G$ and $H$ we write $H=h(G)$ and we say that they are
\emph{isomorphic by} $h$ is $h$ is a bijection of $V_{G}$ onto $V_{H}$\ and
$A_{H}(h(i),h(j))=A_{G}(i,j)$ \ for all $i$ and $j$.\ 

\bigskip

\textit{Pivoting and local complementation}

\textit{\bigskip}

We recall the precise definitions of operations like local complementation and
pivoting, because there are some variants in articles.

For $a$ and $b$ distinct vertices of $G$ we define the graph $H=G^{ab}$ as
follows :

\ $\qquad V_{H}:=V_{G}$ and \ 

$\qquad A_{H}(i,j):=1-A_{G}(i,j)$ \ \ if the following holds :

$\qquad\{i,j\}\cap\{a,b\}=\varnothing$ and

\qquad\qquad\{either \ $i\in N(G,a)-N(G,b)$ and \ $j\in N(G,b),$

\qquad\qquad or \ $j\in N(G,a)-N(G,b)$ and \ $i\in N(G,b),$

\qquad\qquad or \ $i\in N(G,b)-N(G,a)$ and \ $j\in N(G,a),$

\qquad\qquad or \ $j\in N(G,b)-N(G,a)$ and \ $i\in N(G,a)\}.$

\qquad In all other cases, we let $A_{H}(i,j):=A_{G}(i,j).$

\bigskip

This operation is called \emph{pivoting} on $a,b$.\ It does not depend on
whether $a$ and $b$ are loops or are adjacent.

\bigskip

\emph{Local complementation} : for a vertex $a$\ \ of $G$ we define $H=G^{a}$
as follows :

\qquad\ $V_{H}:=V_{G}$ and \ :

$\qquad A_{H}(i,j):=1-A_{G}(i,j)$ \ if $i,j\in N(G,a),$ including the case
$i=j$.

$\qquad A_{H}(i,j):=A_{G}(i,j)$ \ otherwise.

\bigskip

Another notion of local complementation is defined by :

\begin{center}
\ $G\ast a=(G\nabla N(G,a))^{a}=G^{a}\nabla N(G,a).$\ 
\end{center}

It "toggles" the (non loop) edges of $G[N(G,a)]$. It is used for graphs
without loops in the characterization of circle graphs and in the definition
of vertex-minors. ([Bou], [Oum], [CouOum]).

\bigskip

We write $a-b$ to express that $a$ and $b$ are adjacent both without loops,
and $a^{\ell}-b$ to express the same with $a$ looped and $b$ not looped, and
$a^{\ell}-b^{\ell}$ if $a$ and $b$ are looped. These operations satisfy
properties listed in the following lemma :

\bigskip

\textbf{Lemma 1} : For every graph $G=(V,A)$, for distinct vertices $a,b$ and
all sets of vertices $X,Y$ we have :

(1) $(G^{a})^{a}=G$; $G^{ab}=G^{ba}$ ; $(G^{ab})^{ab}=G$ ;

(2) $G^{ab}=h(((G^{a})^{b})^{a}\nabla a)$ if $b\in N(G,a)$ and $h$ is the
permutation of $V$ that exchanges $a$ and $b$ ; $G^{ab}-a-b=((G^{a})^{b}%
)^{a}-a-b.$

(3) $(G^{ab})^{b}=h((G^{a})^{b}\nabla a)$ if $a,b,h$ are as in (2) ;
$(G^{ab})^{b}-a-b=(G^{a})^{b}-a-b.$

(4) $G\nabla X^{ab}=G^{ab}\nabla X$ ; $G\nabla X^{a}=G^{a}\nabla X$ ; $G\nabla
X[Y]=G[Y]\nabla(X\cap Y).$

(5) $G[X]^{ab}=G^{ab}[X]$ ; $G[X]^{a}=G^{a}[X]$\ if $a$ and $b$ are not in
$X.$

\bigskip

\textbf{Proof} : (1), (4), (5) are clear from the definitions.

(2) is a well-known fact about pivoting and local complementation.\ See for
instance [ABS04b].

(3) This is a consequence of (1), (2) and (4) :

($G^{ab})^{b}=(h(((G^{a})^{b})^{a}\nabla a))^{b}$

\qquad\qquad= $h(((G^{a})^{b})^{a}\nabla a)^{a})=h((((G^{a})^{b})^{a}%
)^{a}\nabla a)=h((G^{a})^{b}\nabla a).\square$

\bigskip

\textit{Computating ranks of graphs.}

\bigskip

\textbf{Lemma 2} : For every graph $G$, for distinct vertices $a,b$ we have :

(1) $rk(G)=1+rk(G^{a}-a)$ if $a\in Loops(G)$ ;

(2) $rk(G)=2+rk(G^{ab}-a-b)$ if $a-b$ ;

(3) $rk(G-a)=rk(G^{ab}-a)$ if $a-b;$

(4) $rk(G)=2+rk((G^{a})^{b}-a-b)=1+rk(G^{ab}-b)$ if $a^{\ell}-b$ .

\bigskip

\textbf{Proof }: (1)-(3) are proved in [ABS04b].

(4) We note that $(G^{a})^{b}-a-b=(G^{a}-a)^{b}-b$ and that $(G^{a}-a)$\ \ has
a loop on $b$. Hence by using (1) twice :

\begin{center}
$rk((G^{a})^{b}-a-b)=rk((G^{a}-a)^{b}-b)=rk(G^{a}-a)-1=rk(G)-2.$
\end{center}

For the second equality :

$rk(G)=rk((G^{b})^{a}-a-b)+2=rk((((G^{b})^{a})^{b})^{b}-a-b)+2$

$=rk((((G^{b})^{a})^{b}-a)^{b}-b)+2=rk(((G^{b})^{a})^{b}-a)+1$

because $((G^{b})^{a})^{b}-a$\ \ has a loop on $b$, \ hence :

$rk(G)=rk(h(G^{ba}\nabla b)-a)+1=rk(h(G^{ba}\nabla b-b))+1$

$=rk(G^{ba}\nabla b-b)+1=rk(G^{ab}-b)+1$

because $G^{ba}\nabla b-b=G^{ab}-b.\square$

\bigskip

\section{The multivariate interlace polynomial}

\bigskip

\bigskip

Polynomials have integer coefficients. Following Sokal [Sok]\ \ we call
\emph{multivariate polynomials} those with indeterminates \ $x_{a}$, $y_{a}$,
$z_{a}$,... \ associated with vertices $a$ of the considered graph $G$. We
will denote by $\mathbf{X}_{G}$ the set of such indeterminates for
$X=\{x,y,z,...\}$. They are the \emph{G-indexed indeterminates}. We denote by
$U$ a set $\{u,v,w,...\}$ of "ordinary" indeterminates\ not associated with
elements of graphs.

By a \emph{polynomial} $P(G)$, we mean a mapping $P$ that associates with a
graph $G$ a polynomial in $\mathbf{Z}[U\cup\mathbf{X}_{G}]$ such that if $h$
is an isomorphism of $G$ onto $H$, then $P(H)$ is obtained from $P(G)$ by the
substitution that replaces $x_{a}$ by  $x_{h(a)}$  for every  $x_{a}$ in
$\mathbf{X}_{G}$.

A \emph{specializing substitution} is a substitution that replaces an
indeterminate from a finite set $U=\{u,v,w,...\}$ by a polynomial in
$\mathbf{Z}[U]$, and a $G$-indexed indeterminate $x_{a}$ in $\mathbf{X}_{G}$,
by a polynomial in $\mathbf{Z}[U\cup\{y_{a}\mid y\in X\}]$, the same for each
$a$. For an example, such a substitution can replace $x_{a}$ by $y_{a}%
(x-1)^{2}-3z_{a}u+1,$\ for every vertex $a$ of every graph. If $\sigma$ is a
specializing substitution, then $\sigma\circ P$, defined by \ \ $\sigma\circ
P(G)=\sigma(P(G))$ \ \ is in this sense a polynomial.\bigskip

For a set $A$ of vertices we let $x_{A}$ abbreviate the product (in any order)
of the commutative indeterminates $x_{a}$, for $a$ in $A$. If $A=\varnothing$,
then $x_{A}=1$. If \ $B$ is a set of subsets of $G$, then the polynomial
$\sum\nolimits_{A\in B}x_{A}$ describes exactly $B$. If $B$\ is a multiset of
sets, then it is described by $\sum\nolimits_{A\in B}n(A)\cdot x_{A}$ where
$n(A)$ is the number of occurrences of $A$ in $B$.

\bigskip

\textbf{Definition 3}\ : \textit{The multivariate interlace polynomial.}

\bigskip

For a graph $G$ we define

\begin{center}
$B(G)=\sum\nolimits_{A\cap B=\varnothing}x_{A}y_{B}u^{rk(G\nabla B[A\cup
B])}v^{n(G\nabla B[A\cup B])}$
\end{center}

\ 

\bigskip where $A,B$ range over subsets of $V$.\ \ Hence $B(G)\in
\mathbf{Z}[\{u,v\}\cup\mathbf{X}_{G}\ ]$ where $X=\{x,y\}$.

\bigskip

The interlace polynomial $q$ of of [ABS04b]\ is obtained from $B(G)$\ by a
substitution: $\ q(G;x,y)=\sigma(B(G))$ \ where $\sigma$ \ is the substitution
: \ 

\begin{center}
$[u:=x-1;v:=y-1;x_{a}:=1,y_{a}:=0$ for all $\ a\in V],$
\end{center}

and the polynomial $Q$ of [AvH], defined for graphs without loops is
$Q(G,x)=\tau(B(G))$ \ where $\tau$ \ is the substitution \ 

\begin{center}
$[u:=1;v:=x-2;x_{a}:=y_{a}:=1$ for all $\ a\in V].$

\bigskip
\end{center}

These polynomials are actually \emph{defined recursively} in these articles
(which raises a problem of well-definedness), and then proved to be equal to
the polynomials $\sigma(B(G))$ \ and $\tau(B(G)).$

For more clarity with variable names, we will write $q(G)$ and $Q(G)$ with
variables $u^{\prime}$ and $v^{\prime}$ instead of $x$ and $y$. Hence
\ $\ q(G;u^{\prime},v^{\prime})=\sigma(B(G))$ \ where $\sigma$ \ is the
substitution : \ 

\begin{center}
$[u:=u^{\prime}-1;v:=v^{\prime}-1;x_{a}:=1,y_{a}:=0$ for all $\ a\in V],$
\end{center}

and $Q(G,v^{\prime})=\tau(B(G))$ \ where $\tau$ \ is the substitution \ 

\begin{center}
$[u:=1;v:=v^{\prime}-2;x_{a}:=y_{a}:=1$ for all $\ a\in V].$

\bigskip
\end{center}

Let $B_{1}(G)$ be the polynomial obtained from $B(G)$ by replacing $v$ by 1.\ 

\bigskip

\textbf{Lemma 4}\ : For every graph $G$ and every set $T$ of vertices :

(1) $B(G)=\theta(B_{1}(G))$ where :

$\theta:=[u:=uv^{-1};x_{a}:=vx_{a};y_{a}:=vy_{a}$ for all $\ a\in V],$

(2) $B(G\nabla T)=\mu(B(G))$ where :

$\mu:=[x_{a}:=y_{a},y_{a}:=x_{a}$ for all $\ a\in T].$

\bigskip

\textbf{Proof }: (1) Clear.

(2) We observe that $G\nabla T\nabla B[A\cup B]=G\nabla(A^{\prime}\cup
B^{\prime})[A\cup B]$ where $A^{\prime}=A\cap T,B^{\prime}=B-B\cap T$. The
result follows.$\square$

\bigskip

We will write : $B=\theta\circ B_{1}.$\ \ \ The polynomial $B(G)$ can thus be
"recovered" from $B_{1}(G)$. Since every graph $G$ is $G_{1}\nabla T$ for some
$T$ with $G_{1}$ without loops, we have $B(G)=\mu(B(G_{1}))$ where \ $\mu$\ is
as in Lemma 4. Hence, it is enough to know $B(G)$ for graphs $G$ without
loops.\ However, the recursive definitions to be considered below will
introduce graphs with loops in the recursive calls.

\bigskip

\textit{Properties of polynomials}

\bigskip

The polynomials $q$ and $Q$\ defined above satisfy some properties for all
graphs $G$ :

$\qquad q(G-a)-q(G-a-b)=q(G^{ab}-a)-q(G^{ab}-a-b)$ \ if $\ a-b$(1)

$\qquad Q(G\ast a)=Q(G)$ \ \ \ \ \ \ \ \ \ \ \ \ \ \ \ \ \ \ \ \ \ \ \ \ \ \ \ \ \ \ \ \ \ \ \ \ \ \ \ \ \ \ \ \ \ \ \ \ \ \ \ \ \ \ \ \ \ \ \ \ \ \ \ \ \ \ \ \ \ \ \ \ (2)

$\qquad Q(G^{ab})=Q(G).$ \ \ \ \ \ \ \ \ \ \ \ \ \ \ \ \ \ \ \ \ \ \ \ \ \ \ \ \ \ \ \ \ \ \ \ \ \ \ \ \ \ \ \ \ \ \ \ \ \ \ \ \ \ \ \ \ \ \ \ \ \ \ \ \ \ \ \ \ \ \ \ \ \ \ (3)

Do these equalities hold for $B(G)$ ? The answer is no for (2) and (3) as a
consequence of the next proposition (and also for (1), see below
Counter-example 14).

\bigskip

\textbf{Proposition 5} : A graph $G$ and its polynomial $B(G)$ \ can be
reconstructed from $\rho(B(G))$ where $\rho:=$ $[v:=1;y_{a}:=0$ \ for all
$\ a\in V].$

\bigskip

\textbf{Proof : }For every set of vertices $A$, the rank of $G[A]$ is the
unique integer $n$ such that $x_{A}u^{n}$ is a monomial of $\rho(B(G))$. Now a
vertex $a$ has a loop if $rk(G[a])=1$, and no loop if $rk(G[a])=0$. Hence, we
obtain $Loops(G)$ from $\rho(B(G))$. Using this information, we can
reconstruct edges.

If $a$ and $b$ are not looped, they are adjacent iff \ $rk(G[\{a,b\}]=2$,
otherwise $rk(G[\{a,b\}]=0$.

If one of $a,b$ is looped, they are adjacent iff \ $rk(G[\{a,b\}]=2$,
otherwise $rk(G[\{a,b\}]=1$.

If both are looped, they are adjacent iff \ $rk(G[\{a,b\}]=1$, otherwise
$rk(G[\{a,$ \ \ $b\}]=2$.$\square$

\bigskip

This proof shows how strange is the behaviour of the rank function on graphs.
It follows that identities (2) and (3) cannot hold for $B$ and even for
$\rho\circ B$.

\bigskip

\textbf{Question} : By which algebraic transformations can $B(G)$ be expressed
in terms of $\rho(B(G))$, in a uniform way, valid for all graphs $G$ ?

\bigskip

\subsection{Recursive definition}

\bigskip

\bigskip We now determine a recursive definition, (also called a set of
\emph{reduction formulas}) of $B(G)$ from which will follow the recursive
definitions of [ABS]\ and [AvH].

We let $a$ denote the graph with one non-looped vertex $a$, and $a^{\ell}$ the
similar graph with looped vertex $a$.

\bigskip

\textbf{Lemma 6} : For every graph $G$, for every graph $H$ disjoint from $G$
we have:

(1) $B(\varnothing)=1$

(2) $B(G\oplus H)=B(G)\cdot B(H)$

(3) $B(a)=1+x_{a}v+y_{a}u$

(4) $B(a^{\ell})=1+x_{a}u+y_{a}v.$

\bigskip

\textbf{Proof }: Easy verification from the definitions.$\square$

\bigskip

The more complicated task consists now in expressing $B(G)$ in the case where
$a$ and $b$ are adjacent (this is necessary if no rule of Lemma 6 is
applicable).\ We will distinguish three cases : $a-b$, $a^{\ell}-b$, and
$a^{\ell}-b^{\ell}$.

\bigskip

For a graph $G$ and disjoint sets of vertices $A$ and $B$, we let $m(G,A,B)$
denote the monomial $x_{A}y_{B}u^{rk(G\nabla B[A\cup B])}v^{n(G\nabla B[A\cup
B])}$ so that $B(G)$ is nothing but the sum of them over all pairs $A,B$ (the
condition $A\cap B=\varnothing$ will be assumed for each use of the notation
$m(G,A,B)$).

For distinct vertices $a,b$, two disjoint sets $A,B$ can contain $a,b$ or not
according to 9 cases. We let $i\in\{0,1,2\}$ mean that a vertex is in
$V-(A\cup B)$, in $A$ or in $B$ respectively. Let $B_{ij}$ be the sum of
monomials $m(G,A,B)$ such that $i$ tells where is $a$, and $j$ tells where is
$b$. For an example : $\ B_{02}$ is the sum of monomials $m(G,A,B)$ such that
$a\in V-(A\cup B)$, $b\in B$.

\bigskip

\textbf{Claim 7\ } : Let $G$ with $a-b$.

(1) $B_{00}=B(G-a-b)$

(2) $B_{11}=x_{a}x_{b}u^{2}\cdot B(G^{ab}-a-b).$

(3) $B_{20}=y_{a}u\cdot B(G^{a}-a-b)$ ; \ \ $B_{02}=y_{b}u\cdot B(G^{b}-a-b)$ ;

(4) $B_{12}=x_{a}y_{b}u^{2}\cdot B((G^{b})^{a}-a-b)$ ; \ \ $B_{21}=x_{b}%
y_{a}u^{2}\cdot B((G^{a})^{b}-a-b).$

\bigskip

\textbf{Proof : \ }(1) Clear from the definitions.

(2) A monomial of $B_{11}$ is of the form :

\begin{center}
$m(G,A,B)=x_{A}y_{B}u^{rk(G\nabla B[A\cup B])}v^{n(G\nabla B[A\cup B])}$ \ (1)
\end{center}

with $a,b\in$ $A$ (because of the subscript 11). By Lemma 2(2) we have :

\begin{center}
$rk(G\nabla B[A\cup B])=2+rk(G\nabla B[A\cup B]^{ab}-a-b).$
\end{center}

But $G\nabla B[A\cup B]^{ab}-a-b=(G^{ab}-a-b)\nabla B[A^{\prime}\cup B]$ where
$A^{\prime}=A-a-b$ (we use here Lemma 1(4,5)). Hence :

\begin{center}
$m(G,A,B)=x_{a}x_{b}u^{2}\cdot m(G^{ab}-a-b,A^{\prime},B).$
\end{center}

It follows that $\ :$

\begin{center}
$B_{11}=x_{a}x_{b}u^{2}\cdot B(G^{ab}-a-b)$ \ 
\end{center}

because the set of pairs $A^{\prime},B\subseteq V-a-b$ such that $A^{\prime}$
and $B$ are disjoint coincides with the set of \ pairs $(A-a-b),B$ such that
$A,B\subseteq V$, $A$ and $B$ are disjoint subsets of $V$ and $a,b\in A$.

\bigskip

(3) The proof is similar.\ \ A monomial of $B_{20}$ is of the form (1) above
with $a\in B,$ $b\notin A\cup B$ (because of the subscript 20). By Lemma 2(1)
we have:

\begin{center}
$rk(G\nabla B[A\cup B])=1+rk(G\nabla B[A\cup B]^{a}-a)$
\end{center}

because $a$ is looped in $G\nabla B[A\cup B]$.\ \ But :

\begin{center}
$G\nabla B[A\cup B]^{a}-a=(G\nabla a^{a}-a-b)\nabla B^{\prime}[A\cup
B^{\prime}]$
\end{center}

because $b\notin A\cup B,$ where $B^{\prime}=B-a$. (By Lemma 1). Clearly,
$(G\nabla a^{a}-a-b)=(G^{a}-a-b).$ Hence $m(G,A,B)=y_{a}u\cdot m(G^{a}%
-a-b,A,B^{\prime}).$ It follows that :

\begin{center}
$B_{20}=y_{a}u\cdot B(G^{a}-a-b)$ \ 
\end{center}

because the set of pairs $A,B^{\prime}\subseteq V-a-b$ such that $A$ and
$B^{\prime}$ are disjoint coincides with the set of \ pairs $A,(B-a)$ such
that $A,B\subseteq V$, $A$ and $B$ are disjoint subsets of $V,a\in B$ and
$b\notin A\cup B$. The case of $B_{02}$\ is obtained by exchanging $a$ and $b$.

\bigskip

(4) A monomial of $B_{12}$ is of the form (1) above with $a\in A,$ $b\in B$.
By Lemma 2(4) we have :

\begin{center}
$rk(G\nabla B[A\cup B])=2+rk((G\nabla B[A\cup B]^{b})^{a}-a-b)$
\end{center}

because \ $b^{\ell}-a$ in $G\nabla B[A\cup B]$. We have :

\begin{center}
$(G\nabla B[A\cup B]^{b})^{a}-a-b=((G^{b})^{a}-a-b)$ $\nabla B^{\prime
}[A^{\prime}\cup B^{\prime}]$
\end{center}

where $A^{\prime}=A-a,B^{\prime}=B-b$. Hence :

\begin{center}
$m(G,A,B)=x_{a}y_{b}u^{2}\cdot m((G^{b})^{a}-a-b,A^{\prime},B^{\prime}).$ 
\end{center}

It follows that :

\begin{center}
$B_{12}=x_{a}y_{b}u^{2}\cdot B((G^{b})^{a}-a-b)$ \ 
\end{center}

because the set of pairs $A^{\prime},B^{\prime}\subseteq V-a-b$ such that
$A^{\prime}$ and $B^{\prime}$ are disjoint coincides with the set of \ pairs
$(A-a),(B-b)$ such that $A,B\subseteq V$, $A$ and $B$ are disjoint $a\in A$
and $b\in B$. The case of $B_{21}$\ is obtained similarily by exchanging $a$
and $b$.$\square$

\bigskip

The next claim establishes linear relations between some polynomials $B_{ij}$.

\bigskip

\textbf{Claim 8} : Let $G$ with $a-b$.

(1) $B(G-a)=B_{00}+B_{01}+B_{02}$

(2) $B(G-b)=B_{00}+B_{10}+B_{20}$

(3) $uy_{a}\cdot B(G^{a}-a)=B_{20}+B_{21}+B_{22}$

(4) $uy_{b}\cdot B(G^{b}-b)=B_{02}+B_{12}+B_{22}$

\bigskip

\textbf{Proof }: (1), (2) Clear from the definitions.

(3) From the definitions, \ $B_{20}+B_{21}+B_{22}$\ \ is the sum of monomials
$m(G,A,B)$\ \ such that $a\in B.$ We have :

\begin{center}
$rk(G\nabla B[A\cup B])=1+rk(G\nabla B[A\cup B]^{a}-a)$
\end{center}

by Lemma 2(1). \ But :

$G\nabla B[A\cup B]^{a}-a=((G\nabla a)^{a}-a)\nabla B^{\prime}[A\cup
B^{\prime}]$ \ (where $B^{\prime}=B-a$)

$=(G^{a}-a)\nabla B^{\prime}[A\cup B^{\prime}].$

This gives the result with the usual argument.

(4) Similar to (3) by exchanging $a$ and $b$.$\square$

\bigskip

If we collect the equalities of Claims 7 and 8 we have 10 definitions or
linear equalities for 9 "unknowns". This is enough for obtaining $B(G)$.\ \ We
get thus :

\bigskip

$B(G)=(B_{00}+B_{10}+B_{20})+\{B_{01}+B_{11}+B_{21}\}+(B_{02}+B_{12}+B_{22})$

$=B(G-b)+\{B_{01}+x_{a}x_{b}u^{2}\cdot B(G^{ab}-a-b)+x_{b}y_{a}u^{2}\cdot
B((G^{a})^{b}-a-b)\}+$

$\qquad\qquad+y_{b}u\cdot B(G^{b}-b).$

\bigskip

Then $B_{01}=B(G-a)-B_{00}-B_{02}$

$=B(G-a)-B(G-a-b)-y_{b}u\cdot B(G^{b}-a-b)$ .

\bigskip

We obtain, after reorganization of the expression and a bit of factorization :

\bigskip

\textbf{Lemma 9 \ }: Let $G$ with $a-b$. We have :

$B(G)=x_{b}u^{2}\{x_{a}\cdot B(G^{ab}-a-b)+y_{a}\cdot B((G^{a})^{b}-a-b)\}+$

$\qquad\qquad\qquad\qquad+y_{b}u\{B(G^{b}-b)-B(G^{b}-a-b)\}+$

$\qquad\qquad\qquad\qquad+B(G-a)+B(G-b)-B(G-a-b).$

\bigskip

Considering $B_{22}$\ for which we have two expressions, we get :

\bigskip

\textbf{Corollary 10} : Let $G$ with $a-b$.

$y_{b}\{B(G^{b}-b)-B(G^{b}-a-b)-x_{a}u\cdot B((G^{a})^{b}-a-b)\}$

$=y_{a}\{B(G^{a}-a)-B(G^{a}-a-b))-x_{b}u\cdot B((G^{b})^{a}-a-b)\}.$

\bigskip

Next we consider the cases $a^{\ell}-b$ and $a^{\ell}-b^{\ell}$. Actually,
Lemma 4(2) will shorten the computations.

\bigskip

\textbf{Lemma 11}\ : (1) Let $G$ with $a-b^{\ell}$.

$B(G)=y_{b}u^{2}\{x_{a}\cdot B(G^{ab}-a-b)+y_{a}\cdot B((G^{a})^{b}-a-b)\}$

$\qquad\qquad\qquad\qquad+x_{b}u\{B(G^{b}-b)-B(G^{b}-a-b)\}$

$\qquad\qquad\qquad\qquad+B(G-a)+B(G-b)-B(G-a-b).$

(2) Let $G$ with $a^{\ell}-b^{\ell}$.

$B(G)=y_{b}u^{2}\{y_{a}\cdot B(G^{ab}-a-b)+x_{a}\cdot B((G^{a})^{b}-a-b)\}$

$\qquad\qquad\qquad\qquad+x_{b}u\{B(G^{b}-b)-B(G^{b}-a-b)\}$

$\qquad\qquad\qquad\qquad+B(G-a)+B(G-b)-B(G-a-b).$

\bigskip

\textbf{Proof :} (1) We have $G=G_{1}\nabla b$, $G_{1}=G\nabla b$, where in
$G_{1}$\ we have $a-b$ so that Lemma 9 is applicable.

We get then, letting $\beta$ be the substitution that exchanges $x_{b}$ and
$y_{b}$ :

\bigskip

$B(G)=\beta(B(G_{1}))=$

$=y_{b}u^{2}\{x_{a}\cdot B(G\nabla b^{ab}-a-b)+y_{a}\cdot B((G\nabla
b^{a})^{b}-a-b)\}+$

$\qquad\qquad\qquad\qquad+x_{b}u\{B(G\nabla b^{b}-b)-B(G\nabla b^{b}-a-b)\}+$

$\qquad\qquad\qquad\qquad+\beta(B(G\nabla b-a))+B(G\nabla b-b)-B(G\nabla
b-a-b)$

$=y_{b}u^{2}\{x_{a}\cdot B(G^{ab}-a-b)+y_{a}\cdot B((G^{a})^{b}-a-b)\}$

$\qquad\qquad\qquad\qquad+x_{b}u\{B(G^{b}-b)-B(G^{b}-a-b)\}$

$\qquad\qquad\qquad\qquad+B(G-a)+B(G-b)-B(G-a-b),$

\bigskip

because we have $G\nabla b^{ab}-a-b=G^{ab}-a-b$ and $B(G\nabla b^{ab}-a-b)$
has no occurrence of an indeterminate indexed by $b$,

because we have $(G\nabla b^{a})^{b}-a-b=(G^{a})^{b}-a-b$ and $B((G\nabla
b^{a})^{b}-a-b)$ has no occurrence of an indeterminate indexed by $b$,

and similar remarks apply to $G\nabla b^{b}-b,$ to $G\nabla b^{b}-a-b,$ to
$G\nabla b-b,$ and to $G\nabla b-a-b$.\ Finally, we have $\beta(B(G\nabla
b-a))=B(G-a)$ by Lemma 1 and Lemma 4.

(2) Very similar argument.$\square$

\bigskip

We can now sum up the results of lemmas 6, 9, 11\ into the following
proposition, where the three cases are collected into a single one with help
of the little trick of introducing "metavariables" \ $z_{c},w_{c}$ \ \ for
each $c\in V$ :

$\qquad z_{c}=x_{c}$ and $w_{c}=y_{c}$ if $c$ is not a loop,

$\qquad z_{c}=y_{c}$ and $w_{c}=x_{c}$ if $c$ is a loop.

\bigskip

\bigskip\textbf{Proposition 12} : For every graph $G$, for every graph $H$
disjoint from $G$, every vertex $a$, we have :

(1) $B(\varnothing)=1$

(2) $B(G\oplus H)=B(G)\cdot B(H)$

(3) $B(a)=1+x_{a}v+y_{a}u$

(4) $B(a^{\ell})=1+x_{a}u+y_{a}v$

(5) $B(G)=z_{b}u^{2}\{z_{a}\cdot B(G^{ab}-a-b)+w_{a}\cdot B((G^{a}%
)^{b}-a-b)\}$

$\qquad\qquad\qquad\qquad+w_{b}u\{B(G^{b}-b)-B(G^{b}-a-b)\}$

$\qquad\qquad\qquad\qquad+B(G-a)+B(G-b)-B(G-a-b).$

if $b\in N(G,a)$.

\bigskip

\textbf{Proof : }Immediate consequence of Lemmas 6,9,11. $\square$

\bigskip

We have an even shorter definition :

\bigskip

\textbf{Corollary 13} : For every graph $G$ every vertex $a$, we have :

(1) $B(\varnothing)=1$

(2) $B(G)=(1+z_{a}v+w_{a}u)B(G-a)$ if $N(G,a)=\varnothing$,

(3) $B(G)=z_{b}u^{2}\{z_{a}B(G^{ab}-a-b)+w_{a}B((G^{a})^{b}-a-b)\}$

$\qquad\qquad\qquad\qquad+w_{b}u\{B(G^{b}-b)-B(G^{b}-a-b)\}$

$\qquad\qquad\qquad\qquad+B(G-a)+B(G-b)-B(G-a-b).$

if $b\in N(G,a)$.

\bigskip

\textbf{Counter-example 14 :}

It is proved in \ Proposition 8 \ of [ABS04b] \ that if $a-b$ in $G$ then :

\begin{center}
$q(G-a)-q(G-a-b)=q(G^{ab}-a)-q(G^{ab}-a-b)$ \ \ \ 
\end{center}

\bigskip

This is not true for $B$.\ Take $G=c-a-b-d$. Note that $G^{ab}$ is $G$
augmented with $c-d$. Assume we would have :

\begin{center}
$B(G-a)-B(G-a-b)=B(G^{ab}-a)-B(G^{ab}-a-b).$ \ \ \ (*)
\end{center}

In the left handside, we have a single monomial of the form $y_{b}y_{c}%
x_{d}u^{n}$ for some $n$, and it must be from $B(G-a)$ because $b$ is not in
$G-a-b$. This monomial is $y_{b}y_{c}x_{d}u^{3}$ because $rk(c^{\ell}%
\oplus(d-b^{\ell}))=3$.\ In the right handside we have the monomial
$y_{b}y_{c}x_{d}u^{2}$ because $rk(c^{\ell}-d-b^{\ell})=2$. Hence we cannot
have Equality (*).$\square$

\bigskip

In such a case, we can ask what is the less specialized (or most general)
substitution $\sigma$ such that the corresponding equality is true for
$\sigma\circ B$\ ? Some answers will be given below. We prove actually a more
complicated identity.

\bigskip

\textbf{Proposition 15}\ : If $a-b$ in $G$ then :

\begin{center}
$B(G-a)-B(G-a-b)-B(G^{ab}-a)+B(G^{ab}-a-b)=y_{b}u\{B(G^{b}-a-b)-B((G^{a}%
)^{b}-a-b)\}.$
\end{center}

\bigskip

\textbf{Proof :} We use the notation and some facts from Claims 7 and 8 :

\begin{center}
$B(G-a)-B(G-a-b)=B_{01}+B_{02}=B_{01}+y_{b}u\cdot B(G^{b}-a-b).$
\end{center}

We let $B_{01}^{ab}$ and $B_{02}^{ab}$ denote the polynomials $B_{01}$ and
$B_{02}$ \ relative to $(G^{ab},a,b)$ instead of to $(G,a,b).$ \ Then we have :

\begin{center}
$B(G^{ab}-a)-B(G^{ab}-a-b)=B_{01}^{ab}+B_{02}^{ab}=B_{01}^{ab}+y_{b}u\cdot
B((G^{ab})^{b}-a-b).$
\end{center}

We have by Lemma 1 : $(G^{ab})^{b}-a-b=(G^{a})^{b}-a-b.$

On the other hand, $B_{01}^{ab}$ is the sum of monomials :

\begin{center}
$m(G^{ab},A,B)=x_{A}y_{B}u^{rk(G^{ab}\nabla B[A\cup B])}v^{n(G^{ab}\nabla
B[A\cup B])}$ \ 
\end{center}

for disjoint sets $A,B$ such that $a\notin A\cup B$, $b\in A$. But for such
$A,B$ :

\begin{center}
$G^{ab}\nabla B[A\cup B]=G\nabla B[A\cup B\cup a]^{ab}-a.$
\end{center}

Hence, using Lemma 1 and Lemma 2(3) :

$\qquad rk(G^{ab}\nabla B[A\cup B])=rk(G\nabla B[A\cup B\cup a]^{ab}-a)$

$\qquad=rk(G\nabla B[A\cup B\cup a]-a)$

$\qquad=rk(G\nabla B[A\cup B]).$

\bigskip

We have also $n(G^{ab}\nabla B[A\cup B])=n(G\nabla B[A\cup B]).$ Hence,
$m(G^{ab},A,B)=m(G,A,B)$ and $B_{01}^{ab}=B_{01}$.

Collecting these remarks we get :

$B(G-a)-B(G-a-b)-B(G^{ab}-a)+B(G^{ab}-a-b)$

$=B_{02}-B_{02}^{ab}$

$=y_{b}u\cdot B(G^{b}-a-b)-y_{b}u\cdot B((G^{a})^{b}-a-b).\square$

\bigskip

We note for later use that Identity (*) of Counter-example 14 holds if either
$u=0$ or $y_{b}=0$ for all $b$.

A polynomial $P$ in $\mathbf{Z}[X]$ is said to be \emph{positive} if the
coefficients of its monomials are positive. A polynomial in the sense of a
mapping from graphs to polynomials is \emph{positive} if $P(G)$ is positive
for every graph $G$. It is clear from Definition 3 that $B$ is positive. This
not immediate from the recursive definition of Corollary 13 because of two
substractions in the right handside of (3).\ However, one can derive from
Corollary 13 a stronger statement that is not immediate from Definition 3.

\bigskip

\textbf{Proposition 15 a} : For every graph $G$ and every vertex $a$, the
polynomials $B(G)$ and $B(G)-B(G-a)$ are positive.

\bigskip

\textbf{Proof} : By induction on the number of vertices of $G$, one proves
simultaneously these two assertions by using Corollary 13.

In case (2) we have:

$B(G)-B(G-a)=(1+z_{a}v+w_{a}u)B(G-a)$ and in case (3) we have

$B(G)-B(G-a)=z_{b}u^{2}\{z_{a}\cdot B(G^{ab}-a-b)+w_{a}\cdot B((G^{a}%
)^{b}-a-b)\}$

$\qquad\qquad\qquad\qquad+w_{b}u\{B(G^{b}-b)-B(G^{b}-b-a)\}$

$\qquad\qquad\qquad\qquad+B(G-b)-B(G-b-a),$

which gives with the induction hypothesis that $B(G)-B(G-a)$ is positive.\ So
is $B(G)$ since, again by induction, $B(G-a)$ is positive.$\square$

\bigskip

\section{Specializations to known polynomials}

\bigskip

\bigskip We have already observed that the polynomial $q$ of [ABS04b] \ is
$q(G;u^{\prime},v^{\prime})=\sigma(B(G))$ \ where $\sigma$ \ is the
substitution : \ 

\begin{center}
$[u:=u^{\prime}-1;v:=v^{\prime}-1;x_{a}:=1,y_{a}:=0$ for all $\ a\in V],$
\end{center}

and that the polynomial $Q$ of [AvH], defined for graphs without loops is
$Q(G,v^{\prime})=\tau(B(G))$ \ where $\tau$ \ is the substitution \ 

\begin{center}
$[u:=1;v:=v^{\prime}-2;x_{a}:=y_{a}:=1$ for all $\ a\in V].$
\end{center}

\bigskip

Both are actually specializations of the following two. We let :

$B_{y=0}(G):=\sigma_{0}(B(G))$ where $\sigma_{0}$ \ is the substitution
$[y_{a}:=0$ for all $\ a\in V],$

and

$B_{x=y}(G):=\sigma_{=}(B(G))$ where $\sigma_{=}$ \ is the substitution
$[y_{a}:=x_{a}$ for all $\ a\in V].$

\bigskip

Polynomials $B,B_{x=y},B_{y=0}$ are by definition positive. This is also the
case of $Q$, but this is not obvious from the above definitions. This fact
follows for $Q$ from the recursive definition established in [AvH] that we
will reprove in a different way, but it does not from the definitions given in
[ABS04b].(In the appendix we give a proof that a multivariate version of the
Tutte polynomial is positive.)

\bigskip

\subsection{Fixed loops}

\bigskip

The polynomial $B_{y=0}(G)$ corresponds to "fixed loops" : it does not
describe what happens when some loops are "toggled". It can be rewritten :

\begin{center}
$B_{y=0}(G)=\sum x_{A}u^{rk(G[A])}v^{n(G[A])}.$
\end{center}

Clearly $q(G;u^{\prime},v^{\prime})=\sigma^{\prime}(B_{y=0}(G))$ \ where
$\sigma^{\prime}$ \ is the substitution : \ 

\begin{center}
$[u:=u^{\prime}-1;v:=v^{\prime}-1;x_{a}:=1$ for all $\ a\in V].$
\end{center}

\textbf{Proposition 16} : For every graph $G$ every vertex $a$, we have :

(1) $B_{y=0}(\varnothing)=1$

(2) $B_{y=0}(G)=(1+x_{a}v)B_{y=0}(G-a)$ if $N(G,a)=\varnothing$ and $a$ is not
a loop,

(3) $B_{y=0}(G)=x_{a}u\cdot B_{y=0}(G^{a}-a)+B_{y=0}(G-a)$\ if $a$ is a loop,
isolated or not,

(4) $B_{y=0}(G)=x_{b}x_{a}u^{2}\cdot B_{y=0}(G^{ab}-a-b)+$

$\qquad\qquad+B_{y=0}(G-a)+B_{y=0}(G-b)-B_{y=0}(G-a-b)$ \ \ \ \ if $a-b$.

\bigskip

\textbf{Proof :}\ (1), (2), (4) : Immediate from Corollary 13.

(3) If $a$ is isolated, this follows from Corollary 13 (2). Otherwise, using
the notation of the proof of Claim 7 we note that \ $B_{y=0}(G)$\ is the sum
of monomials $m(G,A,\varnothing)$ ; those such that $a\notin A$ yield
$B(G-a)$, the others yield $x_{a}u\cdot B_{y=0}(G^{a}-a)$ since :

\begin{center}
$rk(G[A])=rk(G[A]^{a}-a)+1=rk((G^{a}-a)[A-a])+1$
\end{center}

by Lemma 2(1). This gives the result, however, it is interesting to see what
gives Lemma 11. The two cases $a^{\ell}-b$ \ and $a^{\ell}-b^{\ell}$ yield the
same equality.

$B_{y=0}(G)=x_{a}u\{B_{y=0}(G^{a}-a)-B_{y=0}(G^{a}-a-b)\}$

$\qquad\qquad\qquad\qquad+B_{y=0}(G-a)+B_{y=0}(G-b)-B_{y=0}(G-a-b).$

Hence we have to check that :

\begin{center}
$x_{a}u\cdot B_{y=0}(G^{a}-a-b)=B_{y=0}(G-b)-B_{y=0}(G-a-b).$
\end{center}

This is nothing but\ Assertion (3) applied to $H=G-b$. Hence (3) can be
established by induction on the size of $G$, with help of Lemma 11, and
without repeating the analysis of the monomials $m(G,A,\varnothing)$.$\square$

\bigskip

This proposition yields, with easy transformations, the following recursive
definition of $q$ :

(q1) $q(G)=v^{\prime n}$ if $G$ consists of $n$ isolated non-looped vertices,

(q2) $q(G)=(u^{\prime}-1)q(G^{a}-a)+q(G-a)$\ if $a$ is a loop, isolated or not,

(q3) $q(G)=(u^{\prime}-1)^{2}q(G^{ab}-a-b)+$

$\qquad\qquad+q(G-a)+q(G-b)-q(G-a-b)$ \ \ \ \ \ \ \ if $a-b$

\bigskip

which is not the one of [ABS04b] Proposition 6. The definition of this article
replaces (q3) by :

(q3') $q(G)=((u^{\prime}-1)^{2}-1)q(G^{ab}-a-b)+$

$\qquad\qquad+q(G-a)+q(G^{ab}-b)$
\ \ \ \ \ \ \ \ \ \ \ \ \ \ \ \ \ \ \ \ \ \ \ \ \ \ \ \ \ \ if $a-b$

\bigskip

However, we have the following corollary of Proposition 15\ that generalizes
Proposition 8 of [ABS04b] :

\bigskip

\textbf{Corollary 17\ : }If $a-b$ in $G$ then :

\begin{center}
$B_{y=0}(G-a)-B_{y=0}(G-a-b)=B_{y=0}(G^{ab}-a)-B_{y=0}(G^{ab}-a-b).$
\end{center}

\bigskip

\textbf{Proof }: Immediate from Proposition 15\ since $y_{b}=0$ for all
$b$.$\ \square$

\bigskip

We get thus the following corollary, and (q3') \ is equivalent to (q3).

\bigskip

\textbf{Corollary 18 } : For every graph $G$ every vertex $a$, we have :

(1) $B_{y=0}(\varnothing)=1$

(2) $B_{y=0}(G)=(1+x_{a}v)B_{y=0}(G-a)$ if $N(G,a)=\varnothing$ and $a$ is not
a loop,

(3) $B_{y=0}(G)=x_{a}u\cdot B_{y=0}(G^{a}-a)+B_{y=0}(G-a)$\ \ \ if $a$ is a
loop, isolated or not,

(4) $B_{y=0}(G)=(x_{b}x_{a}u^{2}-1)B_{y=0}(G^{ab}-a-b)+$

$\qquad\qquad\qquad+B_{y=0}(G-a)+B_{y=0}(G^{ab}-b)$
\ \ \ \ \ \ \ \ \ \ \ \ \ \ \ if $a-b$.

\bigskip

Hence, we have lifted at the multivariate level the recursive definition of
Proposition 6 of [ABS04b].

\bigskip

\textbf{Remark }: The polynomial $q$ is not positive : for $G=a-b$ \ we have
$q(G)=u^{\prime2}-2u^{\prime}+2v^{\prime}.$

\subsection{Mixing loops and nonloops}

\bigskip

We now consider the polynomial $B_{x=y}(G):=\sigma_{=}(B(G))$ where
$\sigma_{=}$ \ is the substitution $[y_{a}:=x_{a}$ for all $\ a\in V].$ \ A
direct (static) definition is :

\bigskip

\begin{center}
$B(G)=\sum\nolimits_{A\cap B=\varnothing}x_{A\cup B}u^{rk(G\nabla B[A\cup
B])}v^{n(G\nabla B[A\cup B])}$
\end{center}

This polynomial has two specializations : first the polynomial $Q$ of [AvH]
defined by $Q(G,v^{\prime})=\tau^{\prime}(B_{x=y}(G))$ \ where $\tau^{\prime}$
\ is the substitution :

\begin{center}
$[u:=1;v:=v^{\prime}-2;x_{a}:=y_{a}:=1$ for all $\ a\in V]$
\end{center}

so that :

\begin{center}
$Q(G,v^{\prime})=\sum\nolimits_{A\cap B=\varnothing}(v^{\prime}-2)^{n(G\nabla
B[A\cup B])}.$

\bigskip
\end{center}

Another one is the \emph{independence polynomial} (Levit and Mandrescu [LM]),
expressible by :

\begin{center}
$I(G,v)=\eta(B_{x=y}(G))$\ 
\end{center}

where $\eta$ \ is the substitution $[u:=0;x_{a}:=1$ for all $\ a\in V].$

\bigskip

\textbf{Proposition 19} : (1) $B_{x=y}(G\nabla T)=B_{x=y}(G)$ for every graph
$G$ and set of vertices $T$.

(2) A graph $G$ without loops and the polynomial $B(G)$ can be uniquely
determined from $\rho(B_{x=y}(G))$, where $\rho$ replaces $v$ by 1.

\bigskip

\textbf{Proof }: (1) follows from Lemma 4.

(2) Consider two distinct vertices $a$ and $b$.\ By looking at the ranks of
the graphs obtained by adding loops to $G[\{a,b\}]$, we see that if $a-b$,
then we have the monomials $x_{a}x_{b}u$ and $3x_{a}x_{b}u^{2}$ in
$\rho(B_{x=y}(G))$. Otherwise, we have the monomials $x_{a}x_{b}$,
$2x_{a}x_{b}u$ and $x_{a}x_{b}u^{2}.\square$

\bigskip

As for Proposition 5, we ask by which algebraic transformation, one can
recover $B(G)$ from $\rho(B_{x=y}(G))$.

\bigskip

Corollary 13 yields the following recursive definition :

\bigskip

\textbf{Proposition 20} : For every graph $G$ :

(1) $B_{x=y}(\varnothing)=1$

(2) $B_{x=y}(G)=(1+x_{a}(u+v))B(G-a)$ \ \ \ \ \ \ \ \ \ \ \ \ \ \ \ \ \ if
$N(G,a)=\varnothing$,

(3) $B_{x=y}(G)=x_{a}x_{b}u^{2}\{B_{x=y}(G^{ab}-a-b)+B_{x=y}((G^{a}%
)^{b}-a-b)\}$

$\qquad\qquad+x_{b}u\{B_{x=y}(G^{b}-b)-B_{x=y}(G^{b}-a-b)\}$

$\qquad\qquad+B_{x=y}(G-a)+B_{x=y}(G-b)-B_{x=y}(G-a-b)$ \ \ \ \ if $b\in
N(G,a)$.

\bigskip

We wish to compare this definition with that presented in [AvH] for $Q$ (and
for graphs without loops) which we recall :

(Q1) $Q(\varnothing)=1$

(Q2) $Q(G)=u^{\prime}\cdot Q(G-a)$
\ \ \ \ \ \ \ \ \ \ \ \ \ \ \ \ \ \ \ \ \ \ \ \ \ \ \ \ \ \ \ \ \ \ if
$N(G,a)=\varnothing$,

(Q3) $Q(G)=Q(G-b)+Q(G\ast b-b)+Q(G^{ab}-a)$ \ \ \ \ \ if $a\in N(G,b)$.

\bigskip

Proposition 20\ yields clearly \ (Q1) and (Q2) but, instead of (Q3) :

(Q3') $Q(G)=Q(G^{ab}-a-b)+Q((G^{a})^{b}-a-b)$

$\qquad\qquad+Q(G^{b}-b)-Q(G^{b}-a-b)$

$\qquad\qquad+Q(G-a)+Q(G-b)-Q(G-a-b)$ \ \ \ \ \ \ \ \ if $b\in N(G,a)$.

\bigskip

However, Proposition 15 yields for $G$ with  $a-b$\ :

$Q(G^{ab}-a)=$

$Q(G-a)-Q(G-a-b)+Q(G^{ab}-a-b)-Q(G^{b}-a-b)+Q((G^{a})^{b}-a-b),$

so that (Q3') reduces to

$Q(G)=Q(G-b)+Q(G^{b}-b)+Q(G^{ab}-a).$

\bigskip

It remains to check that : $Q(G\ast b-b)=Q(G^{b}-b).$ We recall that :

$G\ast b=(G\nabla N(G,b))^{b}=G^{b}\nabla N(G,b).$\ Hence :

$Q(G\ast b-b)=Q(G^{b}\nabla N(G,b)-b)$

$=Q((G^{b}-b)\nabla N(G,b))$

$=Q(G^{b}-b)$

\bigskip

by Proposition 19, as was to be proved. Hence, we have established the
recursive definition of [AvH], but not at the multivariate level. In order to
obtain it from that of Proposition 20, we had to take $u=1$ and $x_{a}=1$ for
all $a$.

The advantage of the definition using (Q1), (Q2), (Q3) is that it only deals
with loop-free graphs, whereas the definition of Proposition 20, even if used
to compute $B_{x=y}(G)$\ for $G$ without loops uses the graphs with loops
$(G^{a})^{b}$\ \ and \ $G^{b}.$ It proves also that $Q$ is positive, which is
not obvious from the definition.

\bigskip

\subsection{The independence polynomial.}

\bigskip

The \emph{independence polynomial} is defined by

\begin{center}
\ \ $I(G,v)=\sum\nolimits_{k}s_{k}v^{k}$
\end{center}

where $s_{k}$ is the number of stable sets of cardinality $k$. (A looped
vertex may belong to a stable set). Hence, we have :

\begin{center}
$I(G,v)=\eta(B_{x=y}(G))$\ 
\end{center}

where $\eta$ \ is the substitution $[u:=0;x_{a}:=1$ for all $\ a\in V].$

\bigskip

We let $B_{I}(G)=\eta^{\prime}(B(G))$\ where $\eta^{\prime}$ \ is the
substitution that replaces $u$ by 0.\ It is a multivariate version of the
independence polynomial, that can be defined directly by :

\begin{center}
$B_{I}(G)=\sum\nolimits_{A,B:\Psi}x_{A}y_{B}v^{n(G\nabla B[A\cup B])}$
\end{center}

where \ $\Psi$ is the set of conditions $A\subseteq V-Loops(G)$, $B\subseteq
Loops(G),G\nabla B[A\cup B]$ has no edge, so that $n(G\nabla B[A\cup B])=\mid
A\cup B\mid.$ From Corollary 13, \ we obtain the recursive definition

(I1) $B_{I}(\varnothing)=1$

(I2) $B_{I}(G)=(1+x_{a}v)B_{I}(G-a)$ \ \ \ \ \ if $N(G,a)=\varnothing$, $a$ is
not a loop,

(I3) $B_{I}(G)=(1+y_{a}v)B_{I}(G-a)$ \ \ \ \ \ \ \ \ if $N(G,a)=\varnothing$,
$a$ is a loop,

(I4) $B_{I}(G)=B_{I}(G-a)+B_{I}(G-b)-B_{I}(G-a-b)$ \ \ \ if $b\in N(G,a)$.

\bigskip

However we can derive alternative reduction formulas :

\bigskip

\textbf{Proposition 21}\ : For every graph $G$ :

(I1) $B_{I}(\varnothing)=1$

(I5) $B_{I}(G)=B_{I}(G-a)+x_{a}v\cdot B_{I}(G-a-N(G,a))$, if $a$ is not a loop,

(I6) $B_{I}(G)=B_{I}(G-a)+y_{a}v\cdot B_{I}(G-a-N(G,a))$, if $a$ is a loop,

(I7) $B_{I}(G)=B_{I}(G-e)-x_{a}x_{b}v^{2}\cdot B_{I}(G-N(G,a)-N(G,b))$, \ \ \ \ \ \ \ \ \ 

\qquad\qquad\qquad\ if $e$ is the edge $a-b$ ; (we do not delete $a$ and $b$
in $G-e$).

\bigskip

\textbf{Proof }: We omit the routine verifications, which use formulas (I1),
(I2), (I3) and induction on size of graphs. $\square$

\bigskip

Formulas (I5), I(6), (I7) are multivariate versions of formulas given in
Proposition 2.1\ of the survey [LM].

\bigskip

\section{Computation of interlace and other monadic second-order polynomials}

\bigskip

We consider whether and how one can \emph{evaluate} for particular values of
indeterminates or \emph{compute} (symbolically) in polynomial time the above
defined multivariate interlace polynomials. The results of this section
reformulate and precise results from [CMR, Mak04, Mak05, Mak06b]. In
particular we consider multivariate polynomials with unbounded numbers of indeterminates.

\bigskip

We define the \emph{size} $\mid P\mid$\ \ of a polynomial $P$\ as the number
of its monomials. Since monomials cannot be written in fixed size, this notion
of size is a lower bound, not an accurate measure of the size in an actual
implementation. However, it is clear that the multivariate polynomial
$B_{y=0}(G)$ has exponential size in the number of vertices of $G$, and so
have $B_{x=y}(G)$ and $B(G)$. Hence, we cannot hope for computing them in
polynomial time., or by edges in other cases) of the considered graphs), hence
as $\mid A\mid+\mid B\mid$ in the case of a monomial $n\cdot x_{A}y_{B}%
u^{p}v^{q}$. For every polynomial $P(G),$\ \ we denote by $P(G)\upharpoonright
d$\ \ its \emph{d-truncation} defined as the sum of its monomials of
quasi-degree at most $d$.

For each $d$, the polynomials $B_{y=0}(G)\upharpoonright d$, $B_{x=y}%
(G)\upharpoonright d$ and $B(G)\upharpoonright d$ have size less than $n^{2d}%
$, where $n$ is the number of vertices. Hence, asking for their computations
in polynomial time has meaning. Since their monomials have integer
coefficients bounded by $n^{2d}$, at most $d$ occurrences of $G$-indexed
indeterminates $x_{a}$, and indeterminates $u,v$ with exponents at most $n$,
we can use their sizes for discussing polynomial time computation of these
truncated polynomials.

For their specializations $P(G)$ where all $G$-indexed indeterminates are
replaced by constants or ordinary indeterminates $x,y,u,...$, we have
$P(G)=P(G)\upharpoonright0$.\ Hence, efficient algorithms for computing
$d$-truncations yield efficient algorithms for computing the classical (non
multivariate) versions of these polynomials, and evaluating them for arbitrary
values of $x,y,u,...$

\bigskip

\textbf{Theorem 22} : For integers $k,d$, for each polynomial $P$ among $B$,
$B_{y=0}$, $B_{x=y}$, $B_{I}$, its $d$-truncation can be computed in time
$O(\mid V\mid^{3d+O(1)})$ for a graph $G$ of clique-width at most $k$.
Polynomials $q(G),Q(G)$ can be computed in times respectively $O(\mid
V\mid^{7})$ and $O(\mid V\mid^{4})$ for graphs of clique-width at most $k$.

\bigskip

We will review clique-width in the next section. This theorem gives for each
$d$ a \emph{fixed parameter tractable algorithm} where clique-width (but not
$d$) is the parameter. See any of the books [DF] and [FG] for the theory of
fixed parameter tractability. As a corollary one obtains the result by
Ellis-Monagham and Sarmiento [ES] that the polynomial $q$ is computable in
polynomial time for \emph{distance-hereditary graphs}, because these graphs
have clique-width at most 3, as proved by Golumbic and Rotics [GolRot].

\bigskip

This theorem will be proved by means of an expression of the considered
polynomials by formulas of monadic second-order logic and for all multivariate
polynomials representable in monadic second-order logic.

\bigskip

\subsection{Clique-width}

\bigskip

\emph{Clique-width} is, like tree-width a graph complexity measure.\ It is
defined and studied in [Cou97, CMR, CouOum, Oum].

\bigskip

Let $C=\{1,...,k\}$ to be used as a set of labels.\ A \emph{k-graph} is a
graph $G$ given with a total mapping from its vertices to $C$,\ denoted by
$lab_{G}$.\ We call $lab_{G}(x)$ the \emph{label} of a vertex $x$. Every graph
is a $k$-graph, with all vertices labelled by 1.

For expressing its properties by logical formulas we will handle a $k$-graph
as a tuple $(V,A,p_{1},...,p_{k})$ where the adjacency matrix $A$ is treated
as a binary relation ($A(x,y)$ is true if $A(x,y)=1$, and false if $A(x,y)=0)$
and $p_{1},...,p_{k}$ are unary relations such that $p_{j}(x)$ is true iff
$lab_{G}(x)=j$.\ 

The operations on $k$-graphs are the following ones :

(i) For each $i\in C$, we define constants \textbf{i} and \textbf{i}$^{\ell}$
for denoting isolated vertices labelled by $i$, the second one with a loop.

(ii) For $i,j\in C$ with $i\neq j$, we define a unary function \ $add_{i,j}%
$\ such that :

$add_{i,j}(V,A,lab)=(V,A^{\prime},lab)$ where $A^{\prime}(x,y)=1$ if
$lab(x)=i$ and $lab(y)=j$ or vice-versa (we want $A^{\prime}$ to be
symmetric), and $A^{\prime}(x,y)=A(x,y)$ otherwise.

This operation adds undirected edges between any two vertices, one labelled by
$i$, the other by $j$, whenever these edges are not already in place.

(iii) We let also $ren_{i\rightarrow j}$ be the unary function such that

$ren_{i\rightarrow j}(V,A,lab)=(V,A,lab^{\prime})$ where $lab^{\prime}(x)=j$
\ if $lab(x)=i$ and $lab^{\prime}(x)=lab(x)$ otherwise. This mapping relabels
by $j$ every vertex labelled by $i.$

(iv) Finally, we use the binary operation $\oplus$\ that makes the union of
disjoint copies of its arguments.\ (Hence $G\oplus G\neq G$ and its number of
vertices is twice that of $G$.)

\bigskip

A well-formed expression \ $t$\ over these symbols will be called a
\emph{k-expression}.\ Its $value$ is a $k$-graph $G=val(t)$. The set of
vertices of $val(t)$ is (or can be defined as) the set of occurrences of the
constants (the symbols \textbf{i} and \textbf{i}$^{\ell}$) in $t$. However, we
will also consider that an expression $t$ designates any graph isomorphic to
$val(t)$. The context specifies whether we consider concrete graphs or graphs
up to isomorphism.

The \emph{clique-width} of a graph $G$, denoted by \emph{cwd(G),} is the
minimal $k$ such that $G=val(t)$ for some $k$-expression $t$. A graph with at
least one edge has clique-width at least 2.\ The graphs $K_{n},S_{n-1}$
($=K_{1,n-1}$) have clique-width 2, for $n\geq3$. It is clear that
clique-width does not depend on loops : $cwd(G\nabla T)=cwd(G)$ for every set
of vertices $T$.

\bigskip

The problem of determining if a graph $G$ has clique-width at most $k$ is
NP-complete if $k$ is part of the input (Fellows et al. [Fell+]).\ However,
for each $k$, there is a cubic algorithm that reports that a graph has
clique-width
$>$
$k$ or produces an \ $f(k)$-expression for some fixed function $f$.\ This
latter result by Oum [Oum] (improved in [HO]) will fit our purposes.\ 

\bigskip

An \emph{ordered} $k$-graph $G$ is a $k$-graph equipped with a linear order
$\leq_{G}$\ on $V$.\ On ordered $k$-graphs, we will use the variant
\ $\overrightarrow{\oplus}$\ \ of $\oplus$\ defined as follows :

\bigskip

(iv) \ $G\overrightarrow{\oplus}\ H$\ is the disjoint union of $G$ and $H$
with a linear order that extends those of $G$ and $H$ and makes the vertices
of $G$ smaller than those of $H$.

The other operations are defined in the same way.

\bigskip

This extension will be used as follows (in 5.3 below) : a graph $G$ being
given with a $k$-expression $t$, we replace everywhere in $t$ the operation
$\oplus$ \ by $\overrightarrow{\oplus}$. The obtained
expression\ $\overrightarrow{t}$\ defines $G$ and a linear ordering on its vertices.

\subsection{Monadic second-order logic}

\bigskip

In a few words, \emph{monadic second-order logic} is \emph{first-order logic}
over powersets.\ Formulas are written with special (uppercase) variables
denoting subsets of the domains of the considered relational structures, and
new atomic formulas of the form $x\in X$ expressing the membership of $x$ in a
set $X$ (and for easier reading, also $x\notin X$). For more details see
[Cou97] and for a use of logic in a field closely related to that of the
present article,\ see [CouOum].

An ordered \ $k$-graph is handled as a relational structure $(V,A,\leq
,p_{1},...,p_{k})$. For a $k$-graph, we simply omit $\leq$. Set variables will
thus denote sets of vertices. Here are some examples of graph properties
expressed in MS logic.

\bigskip

That $G$ is \emph{3-vertex colorable} (with neighbour vertices of different
colors) can be expressed as $G\vDash\gamma$, read "$\gamma$ is true in the
structure $(V,A)$ representing $G$" (here $\leq,p_{1},...,p_{k}$\ do not
matter) : where \ $\gamma$\ \ is the formula :

$\exists X_{1},X_{2},X_{3}\cdot\lbrack\forall x(x\in X_{1}\vee x\in X_{2}\vee
x\in X_{3})\wedge$

$\ \ \ \ \ \ \ \forall x(\lnot(x\in X_{1}\wedge x\in X_{2})\wedge\lnot(x\in
X_{2}\wedge x\in X_{3})\wedge\lnot(x\in X_{1}\wedge x\in X_{3}))$

$\ \ \ \ \ \ \wedge\forall u,v(A(u,v)\wedge\lnot(u=v)\Longrightarrow\lnot(u\in
X_{1}\wedge v\in X_{1})\wedge\lnot(u\in X_{2}\wedge v\in X_{2})$

$\ \ \ \ \ \ \wedge\lnot(u\in X_{3}\wedge v\in X_{3}))].$

\bigskip

\bigskip That $G[B]$ (where $B\subseteq V)$ \ is \emph{not connected} can be
expressed by the formula $\delta(X)$, with $X$ as free variable :

$\exists Y\cdot\lbrack\exists x\cdot(x\in X\wedge x\in Y)\wedge\exists y(y\in
X\wedge y\notin Y)\wedge$

$\qquad\qquad\qquad\forall x,y\cdot(x\in X\wedge y\in X\wedge A(x,y)$

$\qquad\qquad\qquad\qquad\Longrightarrow\{(x\in Y\wedge y\in Y)\vee(x\notin
Y\wedge y\notin Y)\})].$

\bigskip

For $B$ a subset of $V$, ($G,B)\vDash\delta(X)$, read "$\delta$ is true in the
structure representing $G$ with $B$ as value of $X$" \ iff $G[B]$ is not connected.

\bigskip

For building formulas expressing computations in GF(2), we will also use the
set predicate $Even(X)$ expressing that the set denoted by $X$ has even
cardinality. This extended language, called \emph{counting modulo 2} monadic
second-order logic is denoted by C$_{2}$MS.

\bigskip

\textbf{Lemma 23 [CouOum]} : There exists a C$_{2}$MS\ formula $\rho(X,Y)$
expressing that, in a graph $G=(V,A)$\ we have $Y\subseteq X$\ and the row
vectors of $A[X,X]$ associated with $Y$ form a basis of the vector space
defined by the row vectors of the matrix $A[X,X]$. Hence, for each set $X$,
the sets $Y$ satisfying $\rho(X,Y)$ have all the same cardinality, equal to
$rk(G[X])$.

\bigskip

\textbf{Proof} : We first build a basic formula $\lambda(Z,X)$ expressing that
$Z\subseteq X$\ and the row vectors of $A[X,X]$ \ associated with $Z$ are
linearly dependent over GF(2).

Condition $Z\subseteq X$\ is expressed by $\ \forall y\cdot(y\in
Z\Longrightarrow y\in X).$ (We will then use $\subseteq$\ in formulas,
although this relation symbol does not belong to the basic syntax).

The second condition is equivalent to the fact that for each $u\in X$, the
number of vertices $z\in Z$ such that $A(z,u)=1$ is even.\ This fact is
written :

\begin{center}
$\forall u\cdot(u\in X\Longrightarrow\exists W\cdot\lbrack Even(W)\wedge
\forall z\cdot(z\in W\Longleftrightarrow z\in Z\wedge A(z,u))]).$
\end{center}

With $\lambda(Z,X)$ one expresses that $Y$\ (such that $Y\subseteq X)$
\ \ forms a basis by :

\begin{center}
$\lnot\lambda(Y,X)\wedge\forall Z\cdot(\{Y\subseteq Z\wedge Z\subseteq
X\wedge\lnot(Z\subseteq Y)\}\Longrightarrow\lambda(Z,X)).$
\end{center}

We get thus the\ formula $\rho(X,Y)$. $\square$

\bigskip

We will say that the rank function is definable by a C$_{2}$MS-formula.

\bigskip

All basic results (see Section 5.4) hold for\ monadic second-order formulas
written with set predicates $Card_{p}(X)$ expressing that the cardinality of
$X$ is a multiple of $p$ ($Even$ is thus $Card_{2}$).\ This extension of
C$_{2}$MS\ logic called \emph{counting (modulo) monadic second-order, }can be
useful for formalizing computations in fields GF($p$), along the lines of
Lemma 23. For shortness sake we will call \emph{MS\ formula} a formula of this
extended language, (we have no reason to distinguish "pure" monadic
second-order formulas from those using counting modulo predicates), and
\emph{MS\ logic} the corresponding language.

\bigskip

A function $f$ associating a nonnegative integer $f(A,B,C)$\ \ with every
triple of sets $(A,B,C)$ is \emph{defined by an} MS \emph{formula\ \ }%
$\psi(X,Y,Z,U)$ if, for every $(A,B,C)$ the number $f(A,B,C)$ is the common
cardinality of all sets $D$\ such that $(G,A,$ \ $B,C,D)\vDash\psi(X,Y,Z,U)$.
(We distinguish the variables $X,Y,Z,U$ from the sets $A,B,C,D$ they denote).
The generalization to functions $f$ with $k$ arguments is clear, and the
defining formula has then $k+1$ free variables.

\bigskip

\subsection{Multivariate polynomials defined by MS formulas and substitutions}

\bigskip

For an MS \emph{formula\ }$\varphi$ with free variables among $X_{1}%
,...,X_{n}$, for a graph $G$, we let :

\begin{center}
$\mathbf{sat}(G,\varphi,X_{1},...,X_{n})=\{(A_{1},...,A_{n})\mid
A_{1},...,A_{n}\subseteq V,$

$(G,A_{1},...,A_{n})\vDash\varphi(X_{1},...,X_{n})\}$.

\bigskip
\end{center}

This is the set of all $n$-tuples of sets of vertices that satisfy $\varphi$
in $G$.\ We can write it in the form of a multivariate polynomial :

\begin{center}
$P_{\varphi}(G)=\sum\nolimits_{(G,A_{1},...,A_{n})\vDash\varphi(X_{1}%
,...,X_{n})}x_{A_{1}}^{(1)}...x_{A_{n}}^{(n)}$ .
\end{center}

It is clear that \ $P_{\varphi}$\ describes exactly $\mathbf{sat}%
(G,\varphi,X_{1},...,X_{n})$\ and nothing else.\ Its set of indeterminates is
$\mathbf{W}_{G}$ where $W=\{x^{(1)},...,x^{(n)}\}$. The condition describing
the summation will be written in a shorter way $\varphi(A_{1},...,A_{n}).$
Such a polynomial is called a \emph{basic MS polynomial}, and $n$ is its
\emph{order}.

We now show how multivariate polynomials defined by MS\ formulas can be
written as specializations of basic MS-polynomials. To avoid heavy formal
definitions, we consider a typical example :

\begin{center}
$P(G)=\sum\nolimits_{\varphi(A,B,C)}x_{A}y_{B}u^{f(A,B,C)}$ \ \ \ \ \ \ \ \ \ \ \ \ \ \ \ \ \ \ \ \ \ \ \ \ \ \ (0)

\end{center}

where $\varphi(X,Y,Z)$ \ is an MS\ formula and $f$ is a function on triples of
sets defined by an MS formula $\psi(X,Y,Z,U).$\ After usual summation of
\emph{similar} monomials (those with same indeterminates with same exponents)
the general monomial of $P(G)$ is of the form $c\cdot x_{A}y_{B}u^{p}$ where
$c$ is the number of sets $C$ such that $f(A,B,C)=p$. We first observe that
$P(G)=\sigma(P^{\prime}(G))$ where :

\begin{center}
$P^{\prime}(G)=\sum\nolimits_{\varphi(A,B,C)}x_{A}y_{B}z_{C}u^{f(A,B,C)}$
\end{center}

where $\sigma$ replaces each $z_{c}$ by 1. We are looking for an expression of
$P(G)$ as $\mu(\sigma(P_{\theta}(G)))=$ $\mu\circ\sigma(P_{\theta}(G))$
\ \ where $\mu$ replaces each $u_{d}$ by $u$ in :

\begin{center}
$P_{\theta}(G)=\sum\nolimits_{\theta(A,B,C,D)}x_{A}y_{B}z_{C}u_{D}$
\end{center}

for some formula $\theta(X,Y,Z,U).$ Taking $\theta(X,Y,Z,U)$ to be
$\varphi(X,Y,Z)\wedge\psi(X,Y,Z,U)$ would be incorrect in cases where several
sets $D$ satisfy $\psi(A,B,C,D)$ for a triple $(A,B,C)$ satisfying $\varphi.$
We overcome this difficulty in the following way : we let $V$ be linearly
ordered in an arbitrary way ; we let $\psi^{\prime}$ be the formula, written
with a new binary relation symbol denoting the order on $V$ such that
$\ \psi^{\prime}(X,Y,Z,U)$ is equivalent to :

\begin{center}
$\psi(X,Y,Z,U)\wedge\forall T\cdot\lbrack\psi(X,Y,Z,T)\Longrightarrow
"U\leq_{lex}T"]$
\end{center}

where $U\leq_{lex}T$ means that $U$ is less than or equal to $T$ in the
lexicographic order derived from the ordering of $V$.\ This is easily
expressed by an MS formula. The formula $\psi^{\prime}(X,Y,Z,U)$ \ defines the
function $f$ by selecting a unique set $D$\ such that $f(A,B,C)=\mid D\mid$.
This set is unique\ \ for each linear order on $V$ by the hypothesis on
$\psi,$ but its cardinality does not depend on the chosen order.\ Hence we
have the desired expression of $P$ :

\begin{center}
\bigskip$P(G)=\mu\circ\sigma(\sum\nolimits_{\varphi(A,B,C)\wedge\psi^{\prime
}(A,B,C,D)}x_{A}y_{B}z_{C}u_{D}).$
\end{center}

\bigskip These remarks motivate the following definition :

\bigskip\textbf{Definition 23\ a} : \textit{MS-polynomials}

A \emph{multivariate MS-polynomial} is a polynomial of the form :

\begin{center}
$P(G)=\sum\nolimits_{\varphi(A_{1},...,A_{m})}x_{A_{1}}^{(1)}%
...x_{A_{m^{\prime}}}^{(m^{\prime})}u_{1}^{f_{1}(A_{1},...,A_{m})}%
...u_{p}^{f_{p}(A_{1},...,A_{m})}$ \ \ \ \ \ (1)
\end{center}

where $\varphi(X_{1},...,X_{m})$\ is an MS-formula, $m^{\prime}\leq m$ and
$f_{1},...,f_{p}$ \ are MS\ definable functions. A \emph{multivariate
MS-polynomial} in \emph{normal form} is defined as $P=\sigma\circ P_{\varphi}$
\ where :

\begin{center}
$P_{\varphi}(G):=\sum\nolimits_{\varphi(A_{1},...,A_{m})}x_{A_{1}}%
^{(1)}...x_{A_{m}}^{(m)}$ \ \ \ \ \ \ \ (2)

\end{center}

for $\varphi$ an MS-formula and $\sigma$ a substitution that can replace a
variable $x_{a}$ by 1 or by an ordinary variable say $u$. From the above
observations, it is clear that every MS-polynomial can be written in normal
form, for graphs arbitrarily ordered. (The expression of $P$ is said to be
\emph{order-invariant}.)

\bigskip

A \emph{generalized multivariate MS-polynomial} is a polynomial defined as
$P=\sigma\circ P_{\varphi}$ \ where $P_{\varphi}$ \ \ is as in (2) for
$\varphi$ an MS-formula and $\sigma$\ is a specializing substitution. A
generalized MS-polynomial may have negative coefficients, and in this case,
cannot have a normal form.

\bigskip

\textbf{Question 23\ b} : Is it true that every positive generalized
MS-polynomial (i.e., that is positive for every graph) has an expression in
normal form ?

\bigskip

We say that $P$ is of \emph{order} $m$ if it can be expressed as
$P=\sigma\circ P_{\varphi}$ where $\varphi$ has $m$ free variables.\ Hence a
polynomial of the form (1) above is of order $m+p$.

Then for a graph $G$ with $n$ vertices and $P$ defined by (1), $\ P(G)$ has
size at most 2$^{m^{\prime}n}$, degree at most $n(m^{\prime}+p)$, positive
coefficients of value at most 2$^{mn}$. \ These bounds will be useful for
evaluating the cost of computations of truncations of polynomials.

\bigskip

\textit{Transformations of MS-polynomials.}

\bigskip

We now show how some specializations can be reflected by transformations of
the defining formulas. We review some cases which arised in the present
article, by taking a polynomial $P$ of the form (0). \ For each case
1,...,4\ \ we denote by $P_{i}$ the polynomial obtained from $P.$

\bigskip

\textit{Case 1\ }: $x_{a}:=0$. We have :

$P_{1}(G)=\sum\nolimits_{\varphi^{\prime}(B,C)}y_{B}u^{f(\varnothing,B,C)}$

where $\varphi^{\prime}(Y,Z)$ is defined as $\exists X\cdot\lbrack
\varphi(X,Y,Z)\wedge\forall z\cdot z\notin X]$ so that $\varphi^{\prime}(B,C)$
is equivalent to $\varphi(\varnothing,B,C)$. The function $f(\varnothing,Y,Z)$
is defined by $\exists X\cdot\lbrack\psi(X,Y,Z,U)\wedge\forall z\cdot z\notin
X].$

\bigskip

\textit{Case 2} : $u:=0$. We have

$P_{2}(G)=\sum\nolimits_{\varphi^{\prime}(A,B,C)}x_{A}y_{B}$ \ 

where $\varphi^{\prime}(X,Y,Z)$ \ is $\varphi(X,Y,Z)\wedge\exists
U\cdot\lbrack\psi(X,Y,Z,U)\wedge\forall z\cdot z\notin U]$, \ which is
equivalent to $\varphi(X,Y,Z)\wedge"f(X,Y,Z)=0"$.

\bigskip

\textit{Case 3} : $x_{a}:=1$. We have

$P_{3}(G)=\sum\nolimits_{\varphi(A,B,C)}y_{B}u^{f(A,B,C)}$.

The sets $A$ are, like the sets $C$, "invisible" in the monomials. However,
they play a role. They contribute to the multiplicity of the monomials
$y_{B}u^{p}$, for $p=f(A,B,C).$ (This case has been considered above at the
beginning of Section 5.3).

\bigskip

\textit{Case 4} : $x_{a}:=y_{a}$. Here

$P_{4}(G)=\sum\nolimits_{\varphi(A,B,C)}x_{A\cup B}u^{f(A,B,C)}$ \ 

We assume here, and this is enough for this article, that $\varphi
(A,B,C)$\ \ implies that $A$ and $B$ are disjoint. Then to reach the general
syntax we write $P_{4}(G)$ as follows

$P_{4}(G)=\sum\nolimits_{\varphi^{\prime}(D,B,C)}x_{D}u^{g(D,B,C)}$

where $D$ stands for $A\cup B,$ \ $\varphi^{\prime}(W,Y,Z)$ \ is chosen to be
equivalent to the formula $Y\subseteq W\wedge\varphi(W-Y,Y,Z)$ and the
function $g$ is defined by :

$g(D,B,C)=f(D-B,B,C)$

whence also by a formula $\psi^{\prime}(W,Y,Z,U)$ equivalent to :

$\exists X\cdot\lbrack\psi(X,Y,Z,U)\wedge X\subseteq W\wedge"Y=W-X"].$

\bigskip

\textbf{Lemma 24} : Each polynomial $P$ among $B$, $B_{y=0}$, $B_{x=y}$,
$B_{I}$ is an MS-polynomial.\ For every graph $G$, ordered in an arbitrary
way, we have an expression of $P$ in normal form $P(G)=\sigma\circ P_{\theta
}(G)$ for an MS formula $\theta$\ expressing properties of ordered graphs, and
some specializing substitution $\sigma$.

\bigskip

Note that $P(G)=\sigma(P_{\theta}(G))$ for every linear order on $G$.

\bigskip

\textbf{Proof }: We only consider $B$. The other cases follow by the
techniques presented above. We recall the definition of $B$:

\begin{center}
$B(G)=\sum\nolimits_{A\cap B=\varnothing}x_{A}y_{B}u^{rk(G\nabla B[A\cup
B])}v^{n(G\nabla B[A\cup B])}$
\end{center}

\bigskip We let

\begin{center}
$P_{\theta}(G)=\sum\nolimits_{\theta(A,B,C,D)}x_{A}y_{B}u_{C}v_{D}$
\end{center}

where $\theta(A,B,C,D)$ holds iff $A\cap B=\varnothing,C\subseteq A\cup
B,D=A\cup B-C,$ $C$ is the smallest basis of the vector space spanned by
$M_{G\nabla B}[A\cup B,A\cup B],$ where $M_{G\nabla B}$ is the adjacency
matrix of $G\nabla B.$ By "smallest" we mean with respect to the lexicographic
ordering derived from the ordering of $G$. It follows that $\mid
C\mid=rk(G\nabla B[A\cup B])$ and $\mid D\mid=n(G\nabla B[A\cup B]).$ \ By
Lemma 23, one can express these conditions by an MS\ formula $\theta$.

Hence, $B=\sigma\circ P_{\theta}$ \ where $\sigma$ replace $u_{a}$ by $u$ and
$v_{a}$ by $v$.$\square$

\bigskip

\subsection{The Fefermann and Vaught paradigm applied to MS-polynomials}

\bigskip

We now show how multivariate polynomials defined by MS formulas can be
computed "efficiently", not by reduction formulas, because they yield (when
they exist) exponential computations in general, but by induction on
$k$-expressions defining the considered graphs.\ This will only apply to
graphs of clique-width bounded by a fixed value. We will use the
\emph{Fefermann-Vaught paradigm}, presented in detail by Makowsky [Mak04, Mak05].

\bigskip

We need operations that manipulate sets of $q$-tuples, in particular, those of
the form $\mathbf{sat}(G,\varphi,X_{1},...,X_{q}),$ and, equivalently as we
will see, the polynomials $P_{\varphi}(G).$

For sets $R,S$ and $S^{\prime}\subseteq\mathcal{P}(V)^{q},$ \ we write

$R=S\uplus S^{\prime}$ if $R=S\cup S^{\prime}$ and $S\cap S^{\prime
}=\varnothing,$ and

$R=S\boxtimes S^{\prime}$\ if $S\subseteq\mathcal{P}(V_{1})^{q},$
\ $S^{\prime}\subseteq\mathcal{P}(V_{2})^{q},$ $V_{1}\cap V_{2}=\varnothing,$
\ and $R$ \ is the set of $q$-tuples $(A_{1}\cup B_{1},...,A_{q}\cup B_{q})$
such that $(A_{1},...,A_{q})\in S$ and $(B_{1},...,B_{q})\in S^{\prime}.$

\bigskip

For each $S$ $\subseteq\mathcal{P}(V)^{q}$, we let $P(S)$ be the multivariate
polynomial :

\begin{center}
$P(S)=\sum\nolimits_{(A_{1},...,A_{q})\in S}x_{A_{1}}^{(1)}...x_{A_{q}}^{(q)}$ .
\end{center}

Clearly, $P_{\varphi}(G)=P(\mathbf{sat}(G,\varphi,X_{1},...,X_{q})).$ The
following is clear :

\bigskip

\textbf{Fact 25} : For $S,S^{\prime}$ $\subseteq\mathcal{P}(V)^{q}$, we have
$P(S\uplus S^{\prime})=P(S)+P(S^{\prime})$\ and $P(S\boxtimes S^{\prime
})=P(S)\cdot P(S^{\prime}).$

\bigskip

We denote by \textbf{U}$_{k}$ the (finite) set of unary operations allowed in
$k$-expressions. We denote by $MS(k,q)$ the set of MS\ formulas written with
the basic symbols $=,\in,\notin,Card_{p}$ and the relation symbols
$A,<,p_{1},...,p_{k}$ (hence able to express properties of ordered $k$-graphs)
with free variables in the set $\{X_{1},...,X_{q}\}$. The following theorem
discussed in [Mak04] is proved in [Cou97, CouMos] in closely related forms :

\bigskip

\textbf{Theorem 26} : For every $k,q$, for every\ formula $\xi$ in $MS(k,q),$
there exists a finite subset $\Phi$\ of $MS(k,q)$ containing $\xi$ and
satisfying the following properties:

(1) For every $\varphi\in\Phi$ for every $op\in\mathbf{U}_{k}$ there exists a
formula $\varphi^{op}\in\Phi$ such that, for every ordered $k$-graph $G$ :

$\qquad\qquad\qquad\mathbf{sat}(op(G),\varphi,X_{1},...,X_{q})=\mathbf{sat}%
(G,\varphi^{op},X_{1},...,X_{q}).$

(2) For every $\varphi\in\Phi$ there exist $p$ and $(\theta_{1},...,\theta
_{p},\psi_{1},...,\psi_{p})\in\Phi^{2p}$ such that for disjoint ordered
$k$-graphs $G$ and $H$ :

$\qquad\qquad\qquad\mathbf{sat}(G\overrightarrow{\oplus}H,\varphi
,X_{1},...,X_{q})=$

$\qquad\qquad\qquad\qquad{\Large \uplus}_{1\leq i\leq p}\mathbf{sat}%
(G,\theta_{i},X_{1},...,X_{q})\boxtimes\mathbf{sat}(H,\psi_{i},X_{1}%
,...,X_{q}).$

\bigskip

These statements also hold for (unordered) $k$-graphs and the operation
$\oplus$ instead of $\overrightarrow{\oplus}$.

Let $\Phi$ \ be a set of formulas as in Theorem 26. We get a finite family of
polynomials\ $(P_{\varphi})_{\varphi\in\Phi}$ that satisfy mutually recursive
computations rules.\ Actually, the recursive rules apply to the family of
polynomials\ $(\sigma\circ P_{\varphi})_{\varphi\in\Phi}$ where $\sigma$ is a
specializing substitution.We recall that by a polynomial we mean (ambigously)
a mapping $P$ associating with a graph $G$ a polynomial in $\mathbf{Z}%
[U\cup\mathbf{W}_{G}].$

\bigskip

\textbf{Corollary 27} : Let $\Phi$ \ satisfy the properties of Theorem
26.\ Let $\sigma$ be a specializing substitution.\ We have the following
computation rules :

(1) For every $\varphi\in\Phi,$ for every $op\in\mathbf{U}_{k}$ for every
ordered $k$-graph $G$ :

$\qquad\qquad\qquad(\sigma\circ P_{\varphi})(op(G))=(\sigma\circ
P_{\varphi^{op}})(G).$

(2) For every $\varphi\in\Phi$ for every disjoint ordered $k$-graphs $G$ and
$H$ :

$\qquad\qquad\qquad(\sigma\circ P_{\varphi})(G\overrightarrow{\oplus}%
H)=\sum\nolimits_{1\leq i\leq p}(\sigma\circ P_{\theta_{i}})(G)\cdot
(\sigma\circ P_{\psi_{i}})(H).$

where $\varphi^{op}$ and $(\theta_{1},...,\theta_{p},\psi_{1},...,\psi_{p})$
are as in Theorem 26.

\bigskip

\textbf{Proof} : If $\sigma$ is the identity substitution, then (1) and (2)
are direct translations of (1) and (2) of Theorem 26.

Since $\sigma\circ(P+Q)=\sigma\circ P+\sigma\circ Q$ i.e., $(\sigma\circ
P+\sigma\circ Q)(G)=\sigma(P(G))+\sigma(Q(G))=\sigma((P+Q)(G))$\ \ \ and
similarly, \ $\sigma\circ(P\cdot Q)=(\sigma\circ P)\cdot(\sigma\circ Q),$ for
every substitution $\sigma$ and every two polynomials $P,Q$, the equalities
extend to the general case as stated.$\square$

\bigskip

Hence, this corollary concerns all multivariate polynomials described in 5.3.
We will use it for their computation.

\bigskip

\subsection{Computing polynomials in polynomial time}

\bigskip

We discuss the computation of polynomials written in the form $\sigma\circ
P_{\xi}$ \ for $P_{\xi}$ \ a basic MS-polynomial of order $q$, and a
substitution $\sigma$.\ This will also apply to \emph{evaluations }of
polynomials where all indeterminates are given some numeric value, either
integer, real or complex.

The number $n$ will denote the number of vertices of the considered graph $G$.
In evaluating the cost of a computation as $O(e)$ where $e$ is a nonnegative
expression, we omit the case where $e$ might have value 0 (rigourously, we
should write $O(e+1)$). Similarly $\log(e)$ stands for $Max\{1,\log(\mid
e\mid+1)\}$.

Let a graph $G$ be given by a $k$-expression $t$. We "order" $t$ into
$\overrightarrow{t}$ which defines $G$ with a linear order of its vertices
(cf.\ Section 5.1).

It is clear that for each constant, $\mathbf{i}$ or $\mathbf{i}^{\ell}$, each
polynomial $(\sigma\circ P_{\varphi})(\mathbf{i}^{\ell})$ can be computed from
the definitions.\ We can thus compute for each subterm $s$ \ \ of
$\overrightarrow{t}$ \ \ the family of polynomials $((\sigma\circ P_{\varphi
})(val(s)))_{\varphi\in\Phi}.$ In particular, at the end of the computation,
one gets $(\sigma\circ P_{\varphi})(G)$ for all $\varphi\in\Phi.$ (Which is
too much.\ A method for restricting similar computations to their necessary
parts is described in [CouMos] or in [Mak04], Definition 4.17.)

This computation uses at most $n\cdot\mid\Phi\mid$ times the computation rules
of Corollary 27(2), because in a term, the number of occurrence of
\ $\overrightarrow{\oplus}$\ is $s-1$, where $s$ is the number of occurrences
of constants, which is equal to $\mid V\mid=n$. Here, $\mid\Phi\mid$ is
constant. The computation time is bounded by $2n\cdot c_{G}\cdot p_{max}%
\cdot\mid\Phi\mid$, where $p_{max}$ is the maximum value of $p$ in the rules
of Theorem 26(2), and $c_{G}$ bounds the cost of the addition and of the
multiplication of two polynomials. This bound depends on $G$. When do we
obtain a polynomial algorithm in $n?$

\bigskip

We first precise the way we will count the cost of operations on polynomials.
Using \emph{unit cost measure}, we will count for one the cost of each basic
operation on numbers : comparison, addition, substraction, multiplication. The
cost of evaluating $x^{m}$ for a positive integer $m$ is thus $O(\log(m))$. In
the computation of a truncated polynomial $P\upharpoonright d$, we will
consider $d$ as "relatively small" and fixed, like $\mid\Phi\mid$ \ (and
actually much smaller in potential applications.) Hence we will count for one
the cost of computing $x^{m}$ for $m\leq d$.

However, for computing a polynomial or evaluating it for integer values of the
arguments, we can also use the \emph{real cost measure} and consider that the
cost of a comparison, an addition or a substraction of two positive integers
$x$ and $y$ is $O(\log(\mid x\mid+\mid y\mid))$\ and that of their
multiplication is $O(\log(xy))$. Coefficients of polynomials may be
exponentials in the sizes of the considered graphs. However, in situations
where the absolute values of coefficients and exponents are no larger than
2$^{p(n)}$ for fixed polynomials $p$, a polynomial bound on computation time
with respect to unit cost measure remains polynomial with respect to real cost
measure.\ The exponents of the polynomial bounds are just larger.

\bigskip

We will base the following estimations of the cost of computations on
straightforward data structures : a polynomial is a uniquely defined list of
monomials sorted by increasing order of quasi-degree, where two monomials of
same quasi-degree are ordered lexicographically.\ Each monomial is written in
a canonical way by means of a fixed ordering of indeterminates. We deal with
monomials with a variable number of indeterminates, however, this number is
always bounded by $n\cdot\mid X\mid+\mid U\mid$ \ and where $\mathbf{X}_{G}$
is the set of $G$-indexed indeterminates, and $U$ the set of others.

The basic operations on pairs of monomials \ $m,m^{\prime}$ are comparison,
summation of coefficients if \ $m,m^{\prime}$\ are \emph{similar} monomials
(if they have same indeterminates with same respective degrees), and
multiplication. For a monomial $m$, we denote by $\mathbf{v}(m)$ the number of
its indeterminates. The costs of these operations are respectively
$O(\mathbf{v}(m)+\mathbf{v}(m^{\prime})),1,O(\mathbf{v}(m)+\mathbf{v}%
(m^{\prime})).$We denote by $\mathbf{v}(P)$ the number of indeterminates in a
polynomial $P$.\ 

\bigskip

\textbf{Lemma 28 }: (1) For every $P,Q,d$,\ if $\mathbf{v}(P),\mathbf{v}%
(Q)\leq\mathbf{v}_{\max}$ we have:

$(P+Q)\upharpoonright d=P\upharpoonright d+Q\upharpoonright d$ \ and
$(PQ)\upharpoonright d=((P\upharpoonright d)\cdot(Q\upharpoonright
d))\upharpoonright d.$

(2) Computing $P+Q$ \ takes time $\ O(\mathbf{v}_{\max}\cdot(\mid P\mid+\mid
Q\mid))$.

(3) Computing $PQ$ takes time : $O(\mathbf{v}_{\max}\cdot\mid P\mid^{2}%
\cdot\mid Q\mid)$ \ if $\mid P\mid\leq\mid Q\mid.$

\bigskip

\textbf{Proof} : (1) Clear from the definitions.

(2) Note that $\mid P+Q\mid\leq\mid P\mid+\mid Q\mid$. The addition of $P$ and
$Q$ is done by interleaving their lists of monomials and by adding the
coefficients of similar monomials. This gives the result by the remaks on the
cost of operations on monomials.

(3) Let $\mid P\mid\leq\mid Q\mid$ and $\mathbf{v}(P),\mathbf{v}%
(Q)\leq\mathbf{v}_{\max}$. Note that $\mid PQ\mid\leq\mid P\mid\cdot\mid
Q\mid$.\ We compute $PQ$ by multiplying $\mid P\mid$\ times the polynomial $Q$
by a monomial of $P$, and by performing $\mid P\mid-1$ additions of
polynomials of size at most $\mid P\mid\cdot\mid Q\mid$.\ The time taken is at most

\begin{center}
$O(\mathbf{v}_{\max}\cdot\mid P\mid\cdot\mid Q\mid+(\mid P\mid-1)\cdot
\mathbf{v}_{\max}\cdot\mid P\mid\cdot\mid Q\mid)=O(\mathbf{v}_{\max}\cdot\mid
P\mid^{2}\cdot\mid Q\mid).$ $\square$
\end{center}

\bigskip

\textbf{Remark} : If in (3) $P$ and $Q$ are positive, one gets the bound

$O(\mathbf{v}_{\max}\cdot\mid P\mid\cdot\mid PQ\mid)$ because all intermediate
results have size at most $\mid PQ\mid.$

\bigskip

\textbf{Theorem 29} : Let $k,d$ be fixed integers.\ The $d$-truncation of a
generalized\ multivariate MS-polynomial $P(G)$ can be computed in time
$O(n^{6d+O(1)})$ for every graph $G$ of clique-width at most $k$.

\bigskip

The constants hidden by the $O$-notation depend on $P$ and $k$.\ \ The
particular case of evaluations (for numerical values of indeterminates) will
be discussed later. Closely related formulations of this theorem are in [CMR,
Mak04, Mak05]. For a polynomial $P$, we define $\Vert P\Vert:=(\mid P\mid
,\deg(P),C_{\max}(P))$ where $\deg(P)$ is the degree of $P$ and $C_{\max}(P)$
the maximum absolute value of its coefficients. Triples of integers are
ordered componentwise.

\bigskip

\textbf{Lemma \ 30}\ : Let $P$ be a polynomial and $\sigma$ be a substitution
such that \ $\Vert\sigma(x)\Vert\leq(\mathbf{s}_{max},\mathbf{d}%
_{max},\mathbf{c}_{max})$ for every indeterminate $x$. The polynomial
$\sigma\circ P$ satisfies :

\begin{center}
\ \ $\Vert\sigma\circ P\Vert\leq(\mid P\mid\cdot(\mathbf{s}_{max})^{\deg
(P)},\deg(P)\cdot\mathbf{d}_{max},\mid P\mid\cdot C_{\max}(P)\cdot
(\mathbf{s}_{max}\mathbf{c}_{max})^{\deg(P)})$.
\end{center}

\bigskip

\textbf{Proof} : Easy verification. $\square$

\bigskip

It follows from previous observations (see the remark after Question 23 b)
that if \ $P_{\varphi}$ has order $m,$ then for every graph $G,$ the
polynomial $\sigma(P_{\varphi}(G))$ has size and coefficients bounded by
$2^{O(n)\text{ }}$ and degree bounded by $n(m^{\prime}+p)\cdot\mathbf{d}%
_{max}=O(n)$ ($m^{\prime}\leq m$ and $p$ are as in that remark). We can thus
use the unit cost measure.

\bigskip

\textbf{Proof of Theorem 29\ : }Let $k,d$ be integers, let $P$ be a polynomial
expressed as $\sigma\circ P_{\xi}$ \ for an MS\ formula $\xi$\ \ and a
substitution $\sigma$. We aim at computing its $d$-truncation. \ Let $\Phi$ be
the corresponding set of formulas as in Theorem 26. We observe that by Lemma
28 (1), Corollary 27\ yields :

(1) For every $\varphi\in\Phi,$ for every $op\in\mathbf{U}_{k}$ for every
ordered $k$-graph $G$ :

$\qquad\sigma(P_{\varphi}(op(G)))\upharpoonright d=\sigma(P_{\varphi^{op}%
}(G))\upharpoonright d.$

(2) For every $\varphi\in\Phi$ for every disjoint ordered $k$-graphs $G$ and
$H$ :

$\qquad\sigma(P_{\varphi}(G\overrightarrow{\oplus}H))\upharpoonright
d=\sum\nolimits_{1\leq i\leq p}((\sigma(P_{\theta_{i}}(G))\upharpoonright
d)\cdot(\sigma(P_{\psi_{i}}(H))\upharpoonright d))\upharpoonright d.$

\bigskip

Note that we do not have : $\sigma(P(G))\upharpoonright d=\sigma
(P(G)\upharpoonright d)$ in general. As observed above, the time to compute
$\sigma(P(G))$ is $2n\cdot c_{G}\cdot p_{max}\cdot\mid\Phi\mid$, where $c_{G}$
bounds the costs of adding and multiplying the polynomials occuring in
recursion rules (1) and (2). We need only count multiplications which are more
costly than additions and in proportional number. Let us assume that all the
polynomials $\sigma(P_{\varphi}(G))$ are in $\mathbf{Z}[\mathbf{X}_{G}\cup
U]$.\ By Lemma 28\ we have, for\ the $d$-truncations of all such polynomials
$P,Q$\ :

(i) $\ c_{G}=O(\mathbf{v}_{\max}\cdot\mid P\upharpoonright d\mid^{2}\cdot\mid
Q\upharpoonright d\mid)$ with

(ii) $\mathbf{v}_{\max}=n\cdot\mid X\mid+\mid U\mid=O(n),$ and

(iii) $\mid P\upharpoonright d\mid=O(n^{2d+\mid U\mid}).$

For proving (iii), we note that a monomial of $P\upharpoonright d$ is a
product of at most $d$ factors of the form $x_{a}^{s}$ and of at most $\mid
U\mid$ \ factors of the form $u^{s},$ in both cases for $s\leq\deg
(P)\cdot\mathbf{d}_{max}.$ There are \ $n\cdot\mid X\mid\cdot\deg
(P)\cdot\mathbf{d}_{\max}$\ \ \ factors of the form $x_{a}^{s}$ and
$\deg(P)\cdot\mathbf{d}_{\max}$ factors \ $u^{s}$ \ \ for each $u$.\ Hence
\ $\mid P\upharpoonright d\mid=O((n\cdot\deg(P))^{d}\cdot\deg(P)^{\mid U\mid
})$ $=O(n^{2d+\mid U\mid})$ since $\deg(P)=O(n)$\ .

This gives for $2n\cdot c_{G}\cdot p_{max}\cdot\mid\Phi\mid$ the bound
$O(n^{2+6d+3\mid U\mid}).$

We must take into account the cost of building for a graph $G$ of clique-width
at most $k$ a clique-width expression.\ In cubic time, one can construct for
graphs of clique-width at most $k$ an $f(k)$-expression, for a fixed
function\ $f$ ([HO, Oum]). This suffices for our purposes. The total time is
thus $O(n^{t})$ \ where $t=Max\{3,2+6d+3\cdot\mid U\mid\}.$ (This bound
applies if $d=\mid U\mid=0.)$ $\square$

\bigskip

We extend this result to numerical evaluations. Let $P(G)$ be a polynomial in
$\mathbf{Z}[\mathbf{X}_{G}\cup U]$ to be evaluated for a graph $G$.\ \ An
\emph{evaluating substitution}\ $\nu$\ replaces indeterminates by integer,
real or complex values. Let $\nu$ be such a mapping. One can consider
\ $\nu(P(G))$ as a polynomial reduced to a constant, that is the\ desired
value of $P(G)$ for the values of indeterminates specified by $\nu$. Note that
$\nu(P(H))$ is well-defined for every graph $H$ with set of vertices included
in the set $V$ of vertices of $G$. This remark will be useful for the
computation of $\nu(P(G))$ by induction on the structure of $G$ \ using
Corollary 27. The costs of computations are the same for polynomials with
integer, real or complex values since we use unit cost measure.

\bigskip

\textbf{Corollary 31} : Let $k$ be an integer.\ For every
generalized\ multivariate MS-polynomial $P$ and every evaluating substitution,
the corresponding value of $P(G)$ for a graph $G$ of clique-width at most $k$
can be computed in cubic time in the number of vertices of $G$.\ It can be
computed in linear time if the graph is given by a $k$-expression.

\bigskip

\textbf{Proof} : Let $\nu$\ be an evaluating substitution.\ It associates a
number with each $u$ in $U$ and each $x_{a}$ in $\mathbf{X}_{G}$, for a given
graph $G$. As in the proof of Theorem 29\ we have :

(1) For every $\varphi\in\Phi,$ for every $op\in\mathbf{U}_{k}$ for every
ordered $k$-graph $H$ with $V_{H}\subseteq V_{G}$ :

$\qquad\nu(P_{\varphi}(op(H)))=\nu(P_{\varphi^{op}}(H)).$

(2) For every $\varphi\in\Phi$ and all disjoint ordered $k$-graphs $H$ and
$H^{\prime}$ with $V_{H},V_{H^{\prime}}\subseteq V_{G}$ :

$\qquad\nu(P_{\varphi}(H\overrightarrow{\oplus}H^{\prime}))=\sum
\nolimits_{1\leq i\leq p}\nu(P_{\theta_{i}}(H))\cdot\nu(P_{\psi_{i}}%
(H^{\prime})).$

Here we compute values, not polynomials.\ The cost is thus $2n\cdot
p_{max}\cdot\mid\Phi\mid=O(n)$ \ assuming known a $k$-expression defining $G$.
$\square$

\bigskip

We now precise the bounds for the polynomial $B$ and its specializations.
Theorem 22\ is actually a corollary of Theorem 29.

\textbf{Proof of Theorem 22\ : }

For every graph $G$ with $n$ vertices, we have the following bounds :

$\qquad\Vert B(G)\Vert\leq(3^{n},2n,1),\mid B(G)\upharpoonright d\mid
=O(n^{d+2}),$

$\qquad\Vert B_{y=0}(G)\Vert\leq(2^{n},2n,1),\mid B_{y=0}(G)\upharpoonright
d\mid=O(n^{d+1}),$

$\qquad\Vert B_{x=y}(G)\Vert\leq(n2^{n},2n,2^{n}),\mid B_{x=y}%
(G)\upharpoonright d\mid=O(n^{d+2}),$

$\qquad\Vert B_{I}(G)\Vert\leq(2^{n},2n,1)$, $\mid B_{I}(G)\upharpoonright
d\mid=O(n^{d+1}),$

$\qquad\Vert q(G)\Vert\leq(n^{2},2n,2^{n})$,

$\qquad\Vert Q(G)\Vert\leq(n+1,n,3^{n})$.

\bigskip

As in the proof of Theorem 29, we need only bound the costs $c_{G}$ of the
multiplications of polynomials. From these evaluations, we get the bounds
$O(n^{3d+8})$ for \ $B$\ \ and $B_{x=y},O(n^{3d+5}),$ for $B_{y=0}$ and
$B_{I}$, $O(n^{7})$ for \ $q$\ \ and $O(n^{4})$ for $Q$ for the last two,
$\mathbf{v}_{\max}$ is constant. $\square$

\bigskip

\textit{Remark about the size of constants}.

\bigskip

Sets $\Phi$ in Theorem 26 are very large if they are constructed in a blind
manner from an MS formula : the size is a tower of exponentials proportional
to the quantification depth of formula $\xi$. However, if alternatively a
family of polynomials $\sigma\circ P_{\varphi}$ satisfying Corollary 27\ is
constructed directly, by using our knowledge of the meaning of the properties
defined by formula $\xi$, then one may obtain a usable recursive
definition.\ Hence the above estimations leave a great space for improvements.

\bigskip

\section{Conclusion}

\bigskip

We have defined a multivariate interlace polynomial that generalizes the
existing interlace polynomials. The multivariate methodology puts in light the
\emph{meaning of polynomials}.\ Classical polynomials are degraded versions of
multivariate ones. The multivariate approach is well-adapted to the logical
description of polynomials. And the use of monadic second-order\ logic yields
FPT\ algorithms for evaluating polynomials at particular values or for
computing significant portions of them, called truncations.

\bigskip

For computing such polynomials in full, one might use linear delay enumeration
algorithms and try to obtain monomials one by one, by increasing degree, with
a delay between two outputs linear in the size of the next output. Such
algorithms are considered in [Cou06] and [Bag] for MS\ definable problems and
graphs of bounded clique-width.

Another research perspective consists in enriching the notion of
configuration.\ In this article, in particular in Section 5, a configuration
is an $m$-tuple of subsets for fixed $m$. One could try to extend the
methodology of Section 5.3\ to more complex configurations like partitions of
unbounded size or permutations.\ 

\bigskip

Finally, in order to build a zoology as opposed to maintaining a zoo (as
written by J.\ Makowsky [Mak06a]) it is important to relate the various
polynomials by means of algebraic reductions (specializations), or logical
reductions or transformations of other kind.

\bigskip

\textbf{Acknowledgements} : This work is part of a larger project concerning
the logical definition and the complexity of computation of graph polynomials
conducted with J.\ Makowsky. I thank S.\ Oum for the equalities of Lemma 2(4).
Thanks to M.\ Las Vergnas for useful comments.

\bigskip

\section{References}

\bigskip

[ABS] R Arratia, B Bollob\'{a}s, G. Sorkin : The interlace polynomial: a new
graph polynomial.\ J. of Comb.\ Theory \ B\ \textbf{92} (2004) 199-233.

\bigskip

[ABS04b] R Arratia, B Bollob\'{a}s, G. Sorkin : A Two-Variable Interlace
Polynomial, Combinatorica \textbf{24} (2004) 567-584

\bigskip

[And] A. Andrzejak, An algorithm for the Tutte polynomials of graphs of
bounded treewidth, Discrete Mathematics, \textbf{190} (1998) 39-54.

\bigskip

[AvH] M. Aigner, H. van der Holst, Interlace polynomials, Linear algebra and
applications \textbf{377} (2004) 11-30.

\bigskip

[Bag] G. Bagan, MSO\ queries on tree decomposable structures are computable
with linear delay, \textit{Proceedings of Computer Science Logic\ 2006},
Lec.\ Notes Comput.\ Sci. \textbf{4207} (2006) 167-181.

\bigskip

[Bjo] A.\ Bj\"{o}rner, The homology and shellability of matroids and geometric
lattices, Chapter 7 of [Whi].

\bigskip

[Bou] A.\ Bouchet, Circle graph obstructions, Journal of Combinatorial Theory
Series B \textbf{60} (1994) 107-144.

\bigskip

[Cou97] B. Courcelle , The expression of graph properties and graph
transformations in monadic second-order logic, Chapter 5 of the "Handbook of
graph grammars and computing by graph transformations, Vol. 1 : Foundations",
G. Rozenberg ed., World Scientific (New-Jersey, London), 1997, pp. 313-400.

\bigskip

[Cou06] B.\ Courcelle, Linear delay enumeration and monadic second-order
logic, March 2006, to appear in Discrete Applied Mathematics.

\bigskip

[CMR] B. Courcelle, J. A. Makowsky, Udi Rotics: On the fixed parameter
complexity of graph enumeration problems definable in monadic second-order
logic. Discrete Applied Mathematics \textbf{108 }(2001) 23-52

\bigskip

[CGM]\ : B. Courcelle, B. Godlin, J. A. Makowsky : In preparation.

\bigskip

[CouMos] B. Courcelle, M. Mosbah: Monadic Second-Order Evaluations on
Tree-Decomposable Graphs. Theor. Comput. Sci. \textbf{109} (1993) 49-82.

\bigskip

[CouOum] \ B.\ Courcelle, S.\ Oum, Vertex-minors, monadic second-order logic
and a conjecture by Seese, J. comb. Theory B \textbf{97} (2007) 91-126.

\bigskip

[DF] R. Downey et M.\ Fellows, \textit{Parameterized Complexity},
Springer-Verlag, 1999

\bigskip

[ES] J.\ Ellis-Monaghan, I. Sarmiento, Distance Hereditary Graphs and the
Interlace Polynomial, Preprint, arXiv:math.CO/0604088 v1 4 Apr 2006.

\bigskip

[Fell+] M. Fellows, F. Rosamond, U. Rotics, S. Szeider Clique-width
minimization is NP-hard, Proceedings of the thirty-eighth annual ACM symposium
on Theory of computing, Seattle, 2006, pp. 354 - 362.

\bigskip

[FG] \ J.\ Flum, M. Grohe, \textit{Parametrized complexity theory}, Springer, 2006.

\bigskip

[GHN] O.\ Gimenez, P.\ Hlineny, M.\ Noy, Computing the Tutte polynomial on
graphs of bounded clique-width, Procceedings WG\ 2005, Lec.\ Notes
Comput.\ Sci. \textbf{3787} (2005) 59-68.

\bigskip

[GolRot] M.\ Golumbic, U. Rotics: On the Clique-Width of Some Perfect Graph
Classes. Int. J. Found. Comput. Sci. \textbf{11} \ (2000): 423-443

\bigskip

[Hli03] P.\ Hlinen\'{y}, \ On Matroid Properties Definable in the MSO Logic.
Proceedings MFCS 2003, Lec.\ Notes Comp.\ Sci. \textbf{2747} (2003) 470-479

\bigskip

[Hle06] P.\ \ Hlinen\'{y}, The Tutte polynomial for matroids of bounded
branch-width, Combin. Prob. Computing \textbf{15} (2006) 397-409 (Cambridge
Univ. Press).

\bigskip

[HO] P.\ Hlin\v{e}ny, S.\ Oum, Finding branch-decompositions and
rank-decompositions, Preprint, March 2007.

\bigskip

[LV] M.\ Las Vergnas, Le polyn\^{o}me de Martin d'un graphe eul\'{e}rien,
Annals \ of Discrete Mathematics \textbf{17} (1983), 397-411.

\bigskip

[LM] V. E. Levit, E. Mandrescu: The independence polynomials : a survey,
Preprint, october 2005.

\bigskip

[Mak04] J.\ Makowsky : Algorithmic uses of the Feferman--Vaught Theorem,
\ Annals of Pure and Applied Logic \textbf{126} (2004) 159--213

\bigskip

[Mak05] J.\ Makowsky : Coloured Tutte polynomials and Kauffman brackets for
graphs of bounded tree-width, Discrete Applied Maths. \textbf{145} (2005) 276-290.

\bigskip

[Mak06a] J.\ Makowsky : From a Zoo to a Zoology: Descriptive Complexity for
Graph Polynomials, In: A. Beckmann, et al. (eds.): Logical Approaches to
Computational Barriers, Second Conference on Computability in Europe, Lecture
Notes in Computer Science, \textbf{3988} (2006) 330-341.

\bigskip

[Mak06b] J.\ Makowsky, U. Rotics, I. Averbouch and B. Godlin, Computing graph
polynomials on graphs of bounded clique-width, \ In: Proceedings of WG06, H.
Bodlaender et al. (eds), LNCS 4271 (2006) 191-204

\bigskip

[Nob] S. Noble, Evaluating the Tutte polynomial for graphs of bounded
tree-width, Combin. Proba. Computing \textbf{7} (1998) 307-321.

\bigskip

[Oum] \ S. Oum: Approximating Rank-Width and Clique-Width Quickly. Proceedings
WG 2005, LNCS\ \textbf{3787 }(2005) 49-58.

\bigskip

[Sok] A. Sokal: The multivariate Tutte polynomial (alias Potts model) for
graphs and matroids. Surveys in Combinatorics, in Volume \textbf{327}\ of
London Math.\ Soc.\ Lec.\ Notes, 2005, pp. 173-226.

\bigskip

[Whi] N. White (ed.), Matroid applications, Cambridge University Press, 1992.

\bigskip

\section{Appendix : Tutte polynomial}

\bigskip

In this section we apply our logical tools to the Tutte polynomial of
matroids. We define a multivariate Tutte polynomial "better" than Sokal's. We
first establish a lemma.

\bigskip

\subsection{Enumerating polynomials}

The \emph{enumerating polynomial} of a set $S\subseteq\mathcal{P}(V)$ is
defined as $Enum(S)=\sum\nolimits_{A\in S}x_{A}.$ We write $Enum_{x}(S)$ to
specify the generic type $x$ of indeterminates.\ (Hence $Enum_{y}%
(S)=\sum\nolimits_{A\in S}y_{A}).$

For a polynomial $P,$ we let $P^{-}$ (resp. $P^{+}$) denote the polynomial
obtained by replacing each indeterminate $x$ (of any type) by $x-1$ (resp. by
$x+1).$

\bigskip

If $B\subseteq V$, we let $Sub(B):=\{A\mid A\subseteq B\}.$

\bigskip

\textbf{Lemma A.1} : For every $S\subseteq\mathcal{P}(V)$, the following are
equivalent :

(1) $S=Sub(B)$ for some $B\subseteq V$,

(2) $Enum(S)^{-}$ is positive,

(3) $Enum(S)^{-}=x_{B}$ \ for some $B\subseteq V$.

\bigskip

\textbf{Proof }: (1)$\Longrightarrow$ (3) $\Longrightarrow$(2) is
straightforward. The set $B$ in (3) is the same as in (1).

(3) $\Longrightarrow$(1) : We note that $Enum(S)=(Enum(S)^{-})^{+}$ ; if
$Enum(S)^{-}=x_{B}$ then $Enum(S)=x_{B}{}^{+}=Enum(Sub(B))$ \ hence
$S=Sub(B).$

(2)$\Longrightarrow$ (3). By induction on the cardinality of $V$.

We let $S\setminus a$ denote $\{A-a\mid a\in A\in S\}$ and $S-a=\{A\mid
a\notin A\in S\}$. \ We have thus

$\qquad\qquad Enum(S)=x_{a}\cdot Enum(S\setminus a)+Enum(S-a)$, and

$\qquad\qquad Enum(S)^{-}=(x_{a}-1)\cdot Enum(S\setminus a)^{-}+Enum(S-a)^{-}%
.$

It is positive by hypothesis. So is $Enum(S\setminus a)^{-}$\ \ (just look at
monomials with \ $x_{a}$), hence $S\setminus a=Sub(B^{\prime})$ for some
$B^{\prime}\subseteq V-a$.

Then $Enum(S-a)^{-}-Enum(S\setminus a)^{-}$ is also positive.\ Since this is
the case for $Enum(S\setminus a)^{-}$ the same holds for $Enum(S-a)^{-}.$

Hence $S-a=Sub(C)$ for some $C\subseteq V-a$. Hence

$Enum(S)^{-}=(x_{a}-1)x_{B^{\prime}}+x_{C}$. That $Enum(S)^{-}$ is positive
implies $B^{\prime}=C$. Hence $S=Sub(B)$ where $B=B^{\prime}\cup\{a\}$ and
$Enum(S)^{-}=x_{B}$ .$\square$

\bigskip

\subsection{The Tutte polynomial of a matroid}

\bigskip

We represent the Tutte polynomial of a matroid as a specialization of another
multivariate MS-polynomial than the one defined by Sokal in [Sok] recalled in
the introduction. We establish at the multivariate level that its coefficients
are positive, with help of the notion of an \emph{active element} with respect
to a basis and a linear ordering.

\bigskip

We consider the Tutte polynomial for matroids but the application to graphs is
immediate. Our reference is the book edited by N.\ White [Whi]. We refer to
this book for the basic definitions on matroids and some results. A matroid
$M=(E,I)$ with base set $E$ (we can think of $E$ as the edge set of a graph)
is considered as a relational structure where $I$ is a \emph{set predicate}
defining the independent sets. MS\ logic can be used to describe matroids (see
[Hli03], [CouOum]).\ For instance, the bases of $M$\ are characterized by the
MS formula $\beta(X)$ :

\begin{center}
$I(X)\wedge\forall Z\cdot\lbrack\{X\subseteq Z\wedge I(Z)\}\Longrightarrow
Z\subseteq X].$
\end{center}

The \emph{rank polynomial} of $M$ is defined as :

\begin{center}
$R(M)=\sum_{A\subseteq E}x^{r(M)-r(A)}y^{n(A)}$
\end{center}

where $r(M)$ is the rank of $M$, $r(A)$ is that of $A$ and $n(A):=\mid
A\mid-r(A)$. The Tutte polynomial is defined as :

\begin{center}
$T(M):=R(M)^{-}:=\sum_{A\subseteq E}(x-1)^{r(M)-r(A)}(y-1)^{n(A)}$ \ .
\end{center}

Whereas $R(M)$ is clearly positive, this is not obvious for $T(M)$ from this definition.

\bigskip

We let $\sigma$ be the substitution that replaces each $x_{a}$ by $x$ and each
$\ y_{a}$ by $y$. In our logical setting, the rank polynomial can be expressed
as $\sigma\circ\widehat{R}$, for $M$ linearly ordered in an arbitrary way
where :

\begin{center}
$\widehat{R}(M)=\sum_{\varphi(A,C,D)}x_{C}y_{D},$
\end{center}

and the MS-formula$\ $ $\varphi(A,C,D)$ expresses the following conditions :

(a) $A\subseteq E$,

(b) there exists a set $Z$ such that $Z\subseteq A$, $Z$ is a maximal
independent subset of $A$, that is lexicographically minimal with these properties,

(c) $C\subseteq E-A$ is lexicographically minimal such that $Z\cup C$ is a
base of $M$, and

(d) $\ D=A-Z$.

Hence $(C,D,Z)$ is associated in a unique way with $A$ and $\mid
C\mid=r(M)-r(A),$ $\mid D\mid=\mid A\mid-r(A)=n(A).$ It follows that
$R(M)=\sigma(\widehat{R}(M))$.

\bigskip

Hence, the rank polynomial is an MS-polynomial. \ For proving that the Tutte
polynomial is positive, we will replace $\widehat{R}$\ by another multivariate
polynomial, \ in such a way that we can deduce from Lemma A.1\ that the Tutte
polynomial is positive. Our main tool is the following proposition from [Bjo].
We need some definitions relative to an \emph{ordered matroid} $M=(E,I,\leq)$
(i.e., $E$ is linearly ordered).

For every base $B$, an element $e$ of $E-B$ is \emph{externally active with
respect to B} if it is the least element with respect to $\leq$\ of its
\emph{fundamental cycle} (the unique cycle included in $B\cup\{e\}$). Their
set is denoted by $EA(B).$ Dually, an element $b$ of $B$ is \emph{ internally
active in B} if it is the least element of its fundamental cocycle. Their set
is denoted by $IA(B).$

\bigskip

\textbf{Proposition A.2 }([Bjo], Proposition 7.3.6)\ : For every ordered
matroid $M=(E,I,\leq)$ :

\begin{center}
$\mathcal{P}(E)=\uplus_{\beta(B)}\{A\mid B-IA(B)\subseteq A\subseteq B\cup
EA(B)\}.$
\end{center}

\bigskip

We recall that $\uplus$\ denotes a disjoint union and that $\beta
(B)$\ \ expresses that $B$ is a basis.\ Hence every subset $A$ of $E$ can be
decomposed in a \emph{unique way} as :

(i) $A=(A\cap B)\cup(A-B)$

(ii) with $A\cap B\supseteq B-IA(B)$ and $A-B\subseteq EA(B),$ for a unique
base $B$.

(iii) We let $C=B-A$.\ Hence $C\subseteq IA(B).$ \ Then $r(A)=\mid A\cap
B\mid.($See [Bjo] for the proof). It follows that $\mid C\mid=r(M)-r(A).$

(iv) Let $D=A-B$: then $n(A)=\mid D\mid$.

\bigskip The monomial $x_{C}y_{D}$ satisfies :

(v) $\sigma(x_{C}y_{D})=x^{r(M)-r(A)}y^{n(A)}$.

We define :

\begin{center}
$\widetilde{R}(M)=\sum\nolimits_{\beta(B)}\sum_{\theta(B,C,D)}x_{C}y_{D},$
\end{center}

\bigskip where $\theta(B,C,D)$ expresses that $C\subseteq IA(B)$ and
$D\subseteq EA(B)$.

\bigskip\textbf{Claim} : $R(M)=\sigma(\widetilde{R}(M))$.

\textbf{Proof : }Let $A\subseteq E$ and $m(M,A)=x^{r(M)-r(A)}y^{n(A)}$ be the
corresponding monomial in the expression of $R(M)$ \ as $\sum_{A\subseteq
E}m(M,A).$

Let $B$ be the unique basis satisfying conditions (i)-(ii). Let $C$ and $D$ be
associated with $A$ by (iii) and (iv). Then, by (v), $m(M,A)$ is the image by
$\sigma$ of a unique monomial of $\widetilde{R}(M)$.

Conversely, let $x_{C}y_{D}$ be a monomial of $\widetilde{R}(M).$ Let
$A=(B-C)\cup D$. Conditions (i)-(iv) hold, hence $(B,C,D)$ is the unique
triple associated with $A$ by these conditions, and thus $\sigma(x_{C}%
y_{D})=m(M,A)$. This proves the claim.$\square$

\bigskip

Note that for each basis $B$,the condition $\theta(B,C,D)$ expresses that\ $C$
and $D$ can be chosen independently, with $C\subseteq IA(B)$ and $D\subseteq
EA(B)$. Hence $R(M)=\sigma(\widetilde{R}(M))$ where :

\begin{center}

$\widetilde{R}(M)=\sum\nolimits_{\beta(B)}Enum_{x}(Sub(IA(B)))\cdot
Enum_{y}(Sub(EA(B))$.
\end{center}

Let us define

\begin{center}
$\widetilde{T}(M)=\sum\nolimits_{\beta(B)}x_{IA(B)}y_{EA(B)}.$
\end{center}

This is our alternative \emph{multivariate Tutte polynomial for an ordered
matroid} $M$. Clearly,

$\widetilde{R}(M)^{-}=$

$=\{\sum\nolimits_{\beta(B)}Enum_{x}(Sub(IA(B)))\cdot Enum_{y}%
(Sub(EA(B))\}^{-}$

$=\sum\nolimits_{\beta(B)}Enum_{x}(Sub(IA(B)))^{-}\cdot Enum_{y}%
(Sub(EA(B))^{-}$

$=\sum\nolimits_{\beta(B)}Enum_{x}(\{IA(B)\})\cdot Enum_{y}(\{EA(B)\})=\sum
\nolimits_{\beta(B)}x_{IA(B)}y_{EA(B)}$

$=\widetilde{T}(M).$ \ 

We have :

$R(M)^{-}=\sigma(\widetilde{R}(M))^{-}=\sigma(\widetilde{R}(M)^{-}%
)=\sigma(\widetilde{T}(M))=T(M).$

\bigskip

Hence $T(M)$ \ and $\widetilde{T}(M)$ are both positive. We have lifted at the
multivariate level, the well-known identity :

$\qquad\qquad\qquad\qquad T(M)=\sum_{A\subseteq E}(x-1)^{r(M)-r(A)}%
(y-1)^{n(A)}\ $

$\qquad\qquad\qquad\qquad\qquad=\sum\nolimits_{\beta(B)}x^{\mid IA(B)\mid
}y^{\mid EA(B)\mid}$

$\qquad\qquad\qquad\qquad\qquad=R(M)^{-}$

because we have $\ \widetilde{T}(M)=\widetilde{R}(M)^{-}.$

\bigskip

\textit{Computing and evaluating the Tutte polynomial.}

\bigskip

The polynomials

\begin{center}
$\widetilde{R}(M)=\sum\nolimits_{\beta(B)}\sum_{\theta(B,C,D)}x_{C}y_{D},$ and
$\widetilde{T}(M)=\sum\nolimits_{\beta(B)}x_{IA(B)}y_{EA(B)}$
\end{center}

are thus MS-polynomials, because the notion of a basis, the sets $IA(B)$ and
$EA(B)$, the conditions $\theta(B,C,D)$ are MS-expressible. So are $R(M)$ and
$T(M).$ Some consequences regarding their computations follow along the lines
of Theorem 29.

This approach is developped by Makowsky in [Mak05] for graphs of bounded
tree-width. (The application of Theorem 29 is not immediate because a linear
ordering of the edge set must be incorporated to the relational structure in
such a way it remains of bounded tree-width.) Polynomial algorithms that
compute the Tutte polynomial of graphs of bounded tree-width are given in
[And, Nob]; these algorithms do not use monadic second-order logic. However
they use so-called "splitting" formulas which can be seen as particular cases
of equalities (1) and (2) of Theorem 26, which are central for the proof of
Theorem 29 (as pointed out in [Mak05]).

Gimenez et al. give a non-polynomial algorithm that computes the Tutte
polynomial of graphs of bounded clique-width ([GHN]). Theorem 29\ is not
applicable in this case because edge set quantifications are necessary and
MS$_{2}$\ logic does not fit with bounded clique-width.

For matroids of finite branch-width represented on finite fields, Hlineny
gives a polynomial algorithm [Hli06] that follows the ideas of those of [And,
Nob]. However MS properties of such matroids can be checked in polynomial time
(see [Hli03]), hence the method of Theorem 29\ may perhaps be applied also to
these matroids.

\end{document}